\documentclass[preprint,review, 12pt, authoryear]{elsarticle}

\usepackage{amssymb}
\usepackage[hyphens]{url}
\usepackage[breaklinks]{hyperref}   %This is the key package which allows url wrapping.
\usepackage[hyphenbreaks]{breakurl}
\usepackage{longtable}
\usepackage{amsmath,amssymb,multirow,dcolumn,fancyhdr,charter,graphicx}
\usepackage{graphics}
\usepackage{xspace}
\usepackage{color,ulem,epstopdf}
\usepackage[extd]{journalnames} %This is A. Jackson's package that parses ADS journal commands
\usepackage{listings}

\usepackage{setspace}
\begin{document}

\begin{frontmatter}



\title{Automated crater shape retrieval using weakly-supervised deep learning}

 \author[as4,as,as3]{Mohamad Ali-Dib} 
 \author[as2,as5]{Kristen Menou} 
 \author[as,as6]{Alan P. Jackson} 
 \author[as2,as3]{Chenchong Zhu} 
 \author[as]{Noah Hammond}

 \address[as4]{Institut de recherche sur les exoplan\`etes, D\'epartement de physique, Universit\'e de Montr\'eal. 2900 boul. \'Edouard-Montpetit, Montr\'eal, Quebec, H3T 1J4, Canada}     
 
\address[as]{Centre for Planetary Sciences,
    Department of Physical \& Environmental Sciences, University of Toronto Scarborough, Toronto, Ontario M1C 1A4, Canada}
    
\address[as2]{Department of Astronomy \& Astrophysics,
    University of Toronto, Toronto, Ontario M5S 3H4, Canada}
\address[as5]{Physics \& Astrophysics Group, Dept. of Physical \& Environmental Sciences, University of Toronto Scarborough, 1265 Military Trail, Toronto, Ontario, M1C 1A4, Canada}
 \address[as3]{Canadian Institute for Theoretical Astrophysics, University of Toronto, 60 St. George St, Toronto, Ontario M5S 3H8, Canada}   

  \address[as6]{School of Earth and Space Exploration, Arizona State University, 781 E. Terrace Mall, Tempe, AZ 85287, USA}
\end{frontmatter}
Keywords: Moon; Crater Detection; Automation; Deep Learning

\noindent Corresponding authors: 
\begin{itemize}
    \item M.A-D. malidib@astro.umontreal.ca
\end{itemize}

\newpage

\definecolor{APJ}{RGB}{76, 153, 0}
%\textcolor{APJ}{[APJ]}

\section*{Abstract}
Crater {ellipticity} determination is a complex and time consuming task that so far has {evaded successful automation.} We train a state of the art computer vision algorithm to identify craters in Lunar digital elevation maps and retrieve their sizes and {2D shapes.} The computational backbone of the model is MaskRCNN, an ``instance segmentation'' general framework that detects craters in an image while simultaneously producing a mask for each crater that traces its outer rim. Our post-processing pipeline then finds the closest fitting ellipse to these masks, allowing us to retrieve the crater ellipticities. Our model is able to correctly identify 87\% of known craters {in the longitude range we hid from the network during training and validation (test set)}, while predicting thousands of \textcolor{black}{additional} craters not present in our training data.  \textcolor{black}{Manual v}alidation of a subset of these `new' craters indicates that a majority of them are real, which we take as an indicator of the strength of our model in learning to identify craters, despite incomplete training data. The crater size, ellipticity, and depth distributions predicted by our model are consistent with human-generated results. The model allows us to perform a large scale search for differences in crater diameter and shape distributions between the lunar highlands and maria, and we exclude any such differences with a high statistical significance. The predicted test set catalogue and trained model are available  
\href{https://github.com/malidib/Craters_MaskRCNN/}{here}. 

%\end{keyword}

%\begin{linenumbers}
\section{Introduction}
Craters are one of the dominant morphological structures on most solar system objects. Their numbers can be used as a diagnostic tool to estimate the relative surface age of airless objects, while their size distributions hold valuable information on the impactors {(see, for example, reviews by \citealt{werner2015, fassett2016})}. Finding new craters and retrieving their sizes has, however, \textcolor{black}{generally} been a manual process, and as such \textcolor{black}{is} rather time consuming, especially for smaller radii {where the significantly larger number of craters} makes the compilation of global databases difficult.  \textcolor{black}{The impressive database of Martian craters published by \citet{robbins2012}, for example, took years to compile.}

{The usefulness of crater databases, combined with the time consuming nature of manual compilation, means that crater identification and cataloguing has been the target of automation attempts for some time through a variety of methods such as early pattern recognition algorithms \citep{1981PhDT.......182B}, edge-detection \citep[e.g.][]{emami2015}, Hough transforms \citep[e.g.][]{salamuniccar2010}, support-vector machines \citep[e.g.][]{wetzler2005} and decision trees \citep[e.g.][]{stepinski2012}.  Recently attention has turned to neural networks as perhaps the most promising avenue \citep[e.g.][]{cohen2016, palafox2017}.}
A {neural network-based automation} method has been \textcolor{black}{demonstrated} recently by \cite{alidib} who trained a convolutional neural network (CNN) in a UNET architecture (``DeepMoon'') to identify craters on the surface of the Moon with a recall on known craters of 92\%, and a false positive rate on newly identified ones of 11$\pm7$\% (see also the techniques of \cite{harris,christoff}).  

Characterizing the shape and ellipticity distributions of craters, on the other hand, is another hard task that is yet to be decisively automated. Crater ellipticity is controlled {mainly by the impact angle \citep{gault,melosh}, and the surface's geophysical properties}, and is hence useful in order to understand the obliquity history of bodies \citep{holo}, the surface's geophysical properties \citep{elb2,elb1}, and improve age modeling \citep{vesta}. Historically, crater ellipticity was determined through slow and inefficient visual inspection \citep{schultz1982,barlow1988,bottke,herrick}. {It is only relatively recently that numerical methods have been applied to the problem \citep{vesta,robbins2012, robbins2019}, and rarely en masse. }

In this work we present a new method based on MaskRCNN, a neural network capable of identifying craters with accuracy comparable to DeepMoon while simultaneously retrieving their shapes accurately. {This allows for a large sample study of lunar crater ellipticities, and to automate systematic searches for any dichotomies between \textcolor{black}{craters of} the highlands and of the maria.}

We discuss our data sources, training set generation algorithm, methods and neural network architecture in section \ref{method}, present our results on crater size distribution, ellipticity, and depth in section \ref{sec:Results}, and finally conclude in section \ref{conc}.

\section{Methods}
\label{method}

\subsection{Data Preparation}
\label{sec:data}
{All machine learning models needs to be trained on a significant amount of data in order to learn the specific task (in this case crater identification). In particular, convolutional neural networks (CNNs) like MaskRCNN are composed of sequential arrangements of simple non-linear functions - neurons - which have free parameters, and training a CNN means tuning these parameters to control the network's output.  In addition, CNNs have external free hyperparameters such as how many times to pass the input dataset through the network during training (the number of ``training epoch''), and regularization options to prevent overfitting.}

The training data format, however, needs to be consistent with the neural network's architecture.
MaskRCNN works by searching for instances of specific objects (craters) in an image, and returns the bounding boxes containing the object in addition to the associated object masks. This implies that, for each training example, the input needs to be an image of the surface of the Moon containing {one or more} craters, while the training target is a series of independent masks, each covering an individual crater. This data format allows the network to learn to map a lunar surface image into crater masks containing all of the relevant shape information. In practice however, using actual optical images introduces the complex problem of shadows, and so we follow \cite{alidib} in using digital elevation maps (DEMs) instead. This training data format is analogous to the one used by DeepMoon, and hence we use a slightly modified version of the data generation pipeline of \cite{alidib}, \textcolor{black}{which we refer readers to for further details}. This method generates training data by randomly cropping \textcolor{black}{``poststamp''} images \textcolor{black}{representing local areas} from a global digital elevation map \textcolor{black}{of} the Lunar Reconnaisance Orbiter (LRO) and Kaguya merged digital elevation model \citep{barker2016,lolakaguya2015}. This has a {pixel scale} of 512 pixels/degree, or 59 meters/pixel {at the equator}.{ The cropped image is transformed to an orthographic projection using the \texttt{Cartopy Python package} and downsampled to 512 $\times$ 512 pixels.
The location of the cropped image is randomly selected with a uniform distribution, and its size is randomly selected (before downsampling) from a log-uniform distribution with minimum and maximum bounds of 1500 and 6500 pixels (before the second cropping mentioned below). The maximum size was chosen to keep distortion within the postage-stamp images after re-projection to a minimum, while the minimum size was chosen to provide sufficient excess around the edge for the second cropping.} 

\textcolor{black}{The key differences between our data processing pipeline and that of \cite{alidib} are}: 

\begin{itemize}
    \item We downsample the cropped images to $512 \times 512\,\mathrm{pixels}$ instead of $256 \times 256\,\mathrm{pixels}$, as this allows for higher resolution instance masks, leading to a better shape retrieval at the expense of increased memory usage.
    \item {Since our ultimate goal is to retrieve the craters' shapes, we represent the target craters (the CNN's output during the training process) as binary full masks (disks) instead of the binary ``empty'' rings used in DeepMoon. This is to preserve more information on the shapes and facilitate ellipse fitting of the predicted masks in the postprocessing phase (fitting an ellipse to a slightly noisy disk is easier than to a distorted ring), and allows us to be consistent with MaskRCNN architecture that expects full masks. }
    \item Instead of the target being one image containing multiple crater binary rings, each crater now is contained within its own separate target image containing a binary mask. This allows {in principle} for the detection of overlapping craters that would have partially (or completely) disappeared if their binary masks were present in the same image.
    \item We eliminated the padding along the edges of images, as we found these to significantly affect ellipticity retrievals.  This is done by transforming a region about twice the width and height of the desired $512 \times 512\,\mathrm{pixels}$ to an orthographic projection, then cropping the central $512 \times 512\,\mathrm{pixels}$ to obtain a padding-free image. {This means that each input image contains, to varying degrees, a larger portion of lunar surface than those used by \cite{alidib}.}
    \item {In the test set, we separate highlands and maria craters when, for example, calculating the diameter and ellipticity distributions for comparison purposes. However, this separation is only made for the test set, not in the training set, and we use all craters when calculating the recall and precision. }
\end{itemize}

%{Since the largest section of the DEM taken is 6500 pixels across, which at 59 m/pixel is 383.5 km, and taking into account the cropping methodology above, the largest crater our pipeline can detect is $\sim$ 125 km in diameter.}
{Note that the largest DEM section taken by the pipeline is 6500 pixels across, and this corresponds to a physical size of 383.5~km at 59~m/pixel.  This acts as an upper limit to the {diameter} of craters the machine can find, which is reduced further {by a factor 3 (to around 125~km) due to} limitations on the maximum mask size in a given image as described in Section~\ref{sec:postproc}}

The data used to {construct} the targets was obtained by merging the global crater dataset assembled by \cite{povilaitis2017} using the LRO Wide Angle Camera (WAC) Global Lunar DEM, and, for diameters larger than 20 km, the global crater dataset assembled by \cite{head2010} using the LOLA DEM. {Hereafter this combined database is referred to as the P+H database.}

Our training and validation sets span the region of the Moon between -60 and +180 degrees longitude, over the entire latitude range we are considering (-60 to +60 degrees), {while the test set covers the remaining area (where many large lunar maria are present). Note that the lunar maria concentration is approximately centered at 0$^\circ$ longitude, mostly in the northern latitudes.}

We note that during the final phases of writing this manuscript a new global dataset of craters was published by \cite{robbins2019} that includes craters with diameters down to 1-2 km.  {The database compiled by \citet{robbins2019} also contains significantly more craters at sizes larger than 5~km than the P+H database. As discussed by \citet{robbins2019}, this is because they adopt a more liberal/inclusive approach to including surface features that may be craters, whereas \citet{povilaitis2017} and \citet{head2010} adopt a more conservative approach, including only features with a high degree of certainty of being a crater. While this means that the \citet{robbins2019} database may be more complete than the P+H database, it also means that it is likely to have more features {whose interpretation as an impact structure might not be certain}. For machine learning training purposes, it can be better to use a less complete but more conservative database than a more complete one with a higher false positive rate, as the added false positives have the potential to confuse the algorithm.

In addition, the P+H database has the advantage of having been compiled from a DEM like that which we are using for training our model, whereas the \citet{robbins2019} database was compiled using {both optical images and DEMs}.  It is possible that some craters might be clearer in optical images than in the DEM, and vice versa.  For both of these reasons we chose to continue with the P+H database, but note that future authors may wish to try training with the \citet{robbins2019} database, especially if using optical images rather than a DEM.  While we do not re-train the algorithm with the \citet{robbins2019} database, we do use it for comparison, as described in the later sections.

{One caveat of the P+H database is that there is a slight deficit in the number of craters with diameters of around 20~km, at the boundary of the two constituent databases.  \citet{robbins2019} provide a good analysis of this feature in the comparison of their new database to existing ones in the literature (including P+H).  From figure 2 of \citet{robbins2019} we can make a rough estimate that this results in there being 10\% fewer craters with diameters close to 20~km than one would expect.  At this level it is not a dramatic problem, but it is nonetheless something we will return to in some of our later analysis.}
}
%\newpage

%\begin{figure}
%	\begin{center}
%		\centerline{\includegraphics[scale=0.3]{testset.png}}%
%		\caption{Our test set}
%		\label{fig:moondata}
%	\end{center}
%\end{figure}

\subsection{MaskRCNN}

The central algorithm we use to identify craters and retrieve their shapes is MaskRCNN \footnote{\url{https://github.com/facebookresearch/Detectron} and \url{https://github.com/matterport/Mask_RCNN}} (Mask Region Convolutional Neural Network, \cite{rcnn,he}). 

{This versatile framework was designed to detect, within any individual given image, multiple instances for objects of different classes (technically referred to as ``object detection and instance segmentation''). For example, here the aim is to detect multiple craters (the class) in each image. Note that ``under the hood'' MaskRCNN is classifying each pixel, it is hence a ``pixel-wise binary classifier''. In this case, each pixel is classified as either (part of a) ``crater'' or ``background''.}

\begin{figure}
	\begin{center}
		\centerline{\includegraphics[scale=0.3]{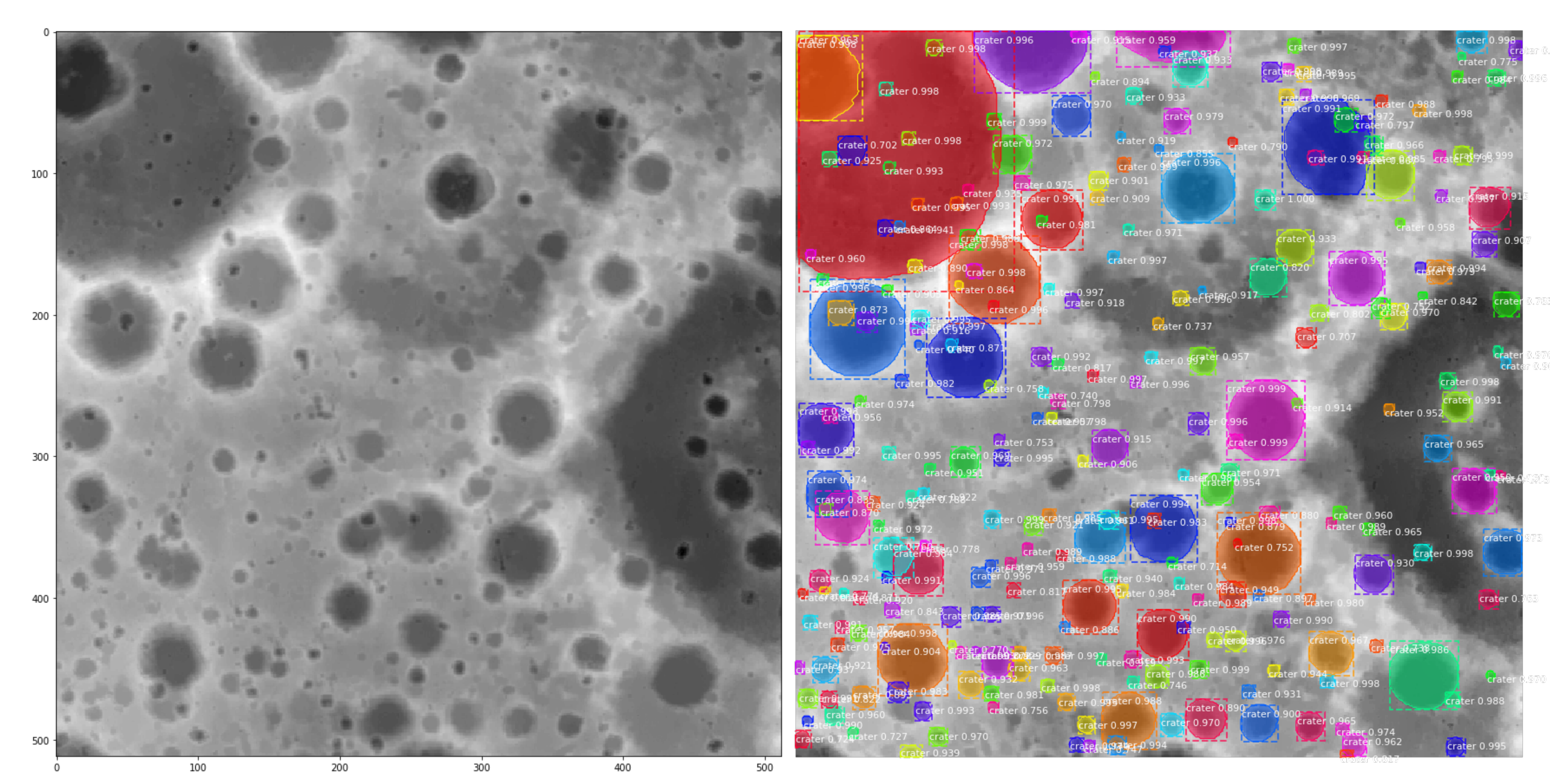}}
		\caption{Poststamp (DEM) image in our test set showing the identified craters bounded by the detection boxes, and their associated shape masks. {The axis labels are pixel numbers.} The numbers shown {near craters} are the detection certainty. This image shows that our algorithm successfully identified tens of craters across a large size spectrum, including overlapping and embedded cases. It also visibly retrieved deviations from circularity in its prediction masks. {Our algorithm correctly identified $\sim$90\% of all catalogued craters in the test set while retrieving size and ellipticity distributions consistent with manual techniques. While its performance decreases for large craters, this is a technical limitation that can be solved with more computational time, and we do not consider it a major issue due to the paucity of these craters. Our algorithm clearly separates overlapping craters in most cases, with a false identification rate due to overlap around 12\% for craters with ellipticities between 1.2 and 2. \textcolor{black}{Craters whose center lies off the edge (there are two good examples at the right-hand edge of this image) are typically not detected in individual poststamps, but this is mitigated by our sampling scheme as they will appear in other poststamps away from the edge.}}}
		\label{fig:detection}
	\end{center}
\end{figure}

{MaskRCNN moreover simultaneously generates a ``segmentation mask'' for each detected instance. Here for example, for each crater it generates a mask covering it and tracing its rim (see Fig. \ref{fig:detection} for examples.). The masks have the same rough shape as the crater, thus giving us shape information. Finally, each mask is contained within a well defined bounding box, giving us spatial information.}

In practice we use the Matterport implementation of MaskRCNN, built with \texttt{Keras} \citet{chollet2015} on top of a \texttt{Tensorflow} \citet{abadi2016a} backend. We modified the default MaskRCNN setup to provide additional augmentation at training time: all images to the network are randomly flipped vertically and horizontally, and they may additionally be rotated by 90, 180 or 270 degrees. {Image augmentation is used to increase the amount and variety of training data.} Additional model configuration details can be found in  Appendix \ref{aa}.

Our MaskRCNN is trained on a dataset of 1980 individual \textcolor{black}{poststamp DEM} images \textcolor{black}{plus their masks}.  An additional 70 images \textcolor{black}{and} masks are used as a validation set. { \textcolor{black}{Each crater has its own individual mask and thus is effectively a single training example.  As the median number of craters per image is $\sim$25 our training set therefore contains $\sim$50,000 training examples.}} All model layers are trained for 80 epochs, using the standard Stochastic Gradient Descent optimizer with a learning rate of $10^{-3}$. The model with the best validation performance on the standard MS COCO\footnote{http://cocodataset.org} mAP50 metric (mean Average Precision at an intersection over union, or Jaccard Index, of 0.5) is kept for predictions on the test set. In essence this is a maximization of mask overlap that also penalizes against false detections.

{Model weights are initialized with ``MS COCO'' pretrained values, so our training procedure can be seen as effectively fine-tuning a pretrained CNN to our specific dataset. This means that we took a model that is already capable (through earlier training) of detecting and classifying everyday objects in an image (people, cars, etc..), and that already ``knows'' fundamental concepts such as edges and natural world textures. By starting from this pretrained model, it is easier for the CNN to learn more complicated concepts such as crater rims. }

Grey DEM images are fed to MaskRCNN as identically valued in each of the three RGB channels of the model. For better contrast, the images are pre-processed with the Contrast Limited Adaptive Histogram Equalization (CLAHE) algorithm, as implemented in the \texttt{OpenCV} library. In practice, we apply this contrast enhancement \textcolor{black}{only on lightness (L-band)} after an RGB to LAB color space transformation (see Appendix \ref{ab} for the detailed preprocessing algorithm). \textcolor{black}{Note that prior to any other processing the source DEM is downsampled from 16-bit greyscale to 8-bit greyscale, as in the DeepMoon pipeline. {This is to allow for the training set to fit on the GPU memory. We emphasize, however, that 8-bit topography could wash out shallow impact craters which could significantly affect their detection, especially in cases when superposed impactors may display fairly subtle topography.}} {Note that to calculate the craters depth (section 3.4), we use the original unprocessed DEMs where the pixel values are linear to elevation.}

\subsection{Post-Processing}
\label{sec:postproc}
After visually inspecting our test set predictions, we found several (but rare) instances where the machine failed entirely, predicting a mask that covers almost the entire poststamp image. This is likely due to the complexity of the machine learning task of predicting the same object at very different size scales. We hence filter all masks whose pixel area is bigger than 1/3 of a 512x512 image. This corresponds to roughly 0.3\% of the total unique craters. {Combined with the image processing pipeline described in Section~\ref{sec:data}, this cut means that the largest crater the machine can detect is $\sim$125~km in diameter.  As larger craters are small in number and easily found by hand we do not consider this a major shortcoming.}

While MaskRCNN predicts the position and shape of a crater through respectively the bounding box and mask it returns, we still need to retrieve the physical diameter and ellipticity from the boxed mask. 
Calculating the diameter is a straightforward task, as to first order it suffices to define the diameter (in pixels) as simply the length of the mask's bounding box, and then (knowing the poststamp image's location on the Moon) convert this quantity to real diameter in kilometers. {In practice we use the average of the two (x, y) lengths.} Any errors introduced by this method are small compared to the uncertainties we have on the measured crater diameters. Therefore, we are virtually getting the crater diameters for ``free'', compared to DeepMoon where a separate pipeline had to be implemented in order to \textcolor{black}{calculate these values with a brute-force fitting algorithm}. {Note that we ignore any errors on the diameters introduced by the ellipticity of the craters resulting in some bounding boxes deviating from exact squares. This is justified by the relatively low number of highly elliptical craters.}

Retrieving the ellipticity on the other hand is formally a computer vision task consisting of fitting the near-circular mask image to an ellipse template and then measuring its semi-major and semi-minor axes. We do this using using \texttt{OpenCV}'s ``\texttt{fitEllipse}'' module \citep{opencv_library} implemented using the direct least-square fitting method of \cite{ellipse}. This module takes the crater mask as input, then returns the center of the fitted ellipse, its major and minor axes, and an orientation angle. {Individual predicted masks in MaskRCNN are returned as binary images, and hence the boundary of the crater is traced by the outermost non-zero pixels.} We finally define the ellipticity ($\epsilon$) of the crater as the ratio of its major to minor axis. {We note that the direct least-squares fitting of ellipses has well documented biases, and some other methods such as \cite{szpak2015} can perform better under some circumstances \citep{robbins2019}. This is discussed further in section \ref{sec:ellrobbins}.}

{We emphasise that we do not have a ground truth for crater ellipticity from the P+H catalogues (even though \cite{robbins2019} provided one that we compare against later), the ground truth masks on which MaskRCNN is trained are all perfectly circular.} Therefore the machine learns the non-circular shape deviations as a byproduct of detecting craters with various shapes, something referred to as ``weak-supervision''. In other words, we do not directly teach the machine to identify shapes, it learns them as part of its primary detection-focused task. {The main value of this approach is that the algorithm learns the shape without any prior bias on said shape. It could have failed entirely (for example, returning perfect circle masks, as seen in training, for all, even non-circular craters). Instead we find that it does return elliptical shapes, implying that the model did indeed learn this concept.}

The results of section 3.3 should thus be taken with this in mind and viewed more as a guideline of what can be achieved by machine learning algorithms.  The comparison between ellipticity data obtained by MaskRCNN and literature studies also offers a test on how well the algorithm is learning to identify the features that define the shape of a crater.

Note that the number of craters with significant ellipticity is expected to be small so this is unlikely to have a noticeable effect on the basic task of crater identification.

\textcolor{black}{After we determine crater properties for each poststamp image, additional calculations are required to derive a global crater catalog from them.}  Since we \textcolor{black}{randomly crop poststamp images from a single global DEM} to generate our dataset, the same crater can (and usually does) appear in multiple \textcolor{black}{images}. To generate accurate \textcolor{black}{global} crater size and shape distributions, we need to filter these duplicate craters. We hence use the same method as \cite{alidib} (section 2.5), where we classify craters as duplicates if their \textcolor{black}{longitudes $\mathcal{L}$, latitudes $L$, and radii $R$ overlap within a certain tolerance factor}:

\begin{equation}
\label{eq1}
    {\frac{\left(\left(\mathcal{L}_{i}-\mathcal{L}_{j}\right)^{2} \cos ^{2}\left(\frac{\pi}{180^{\circ}}\langle L\rangle\right)+\left(L_{i}-L_{j}\right)^{2}\right)}{C_{K D}^{2} \min \left(R_{i}, R_{j}\right)^{2}}<D_{\mathcal{L}, L}}
    \end{equation}

    \begin{equation}
    {\frac{\operatorname{abs}\left(R_{i}-R_{j}\right)}{\min \left(R_{i}, R_{j}\right)}<D_{R}}
\end{equation}

with $C_{K D}=\frac{180^{\circ}}{\pi R_{\mathrm{Moon}}}$ and $\langle L\rangle=\frac{1}{2}\left(L_{i}+L_{j}\right)$.
{Note that eq. \ref{eq1} is mathematically equivalent to the more intuitive form $$\left(\left(x_{i}-x_{j}\right)^{2}+\left(y_{i}-y_{j}\right)^{2}\right) / \min \left(r_{i}, r_{j}\right)^{2}<D_{x, y}$$ but where we converted the (x, y) coordinates into absolute (longitude, latitude). Here we use $D_{\mathcal{L}, L}$=2.08 and D$_{R}$=1.44, values we found (through trial-and-error) to minimize the false-duplicate error rate. 
To assess this approach and check its validity we created a set of 300 randomly chosen pairs of craters classified as duplicates by the algorithm using these parameters, then visually inspected them. We estimated the false-duplicate error rate to be around $\sim 4\%$, with craters {nested} within other craters of comparable diameter being its main source. Since the sample was taken from the entirety of the test set, this number should be considered as a number-weighted average between highland and maria craters.}

\section{Results}
\label{sec:Results}
\subsection{Crater Identification on the Moon}
\label{sec:iden}
The first and most basic task our model needs to do is crater identification: in any given DEM, find all existing craters. The performance of such machines is measured via precision and recall, defined respectively as: \\
\begin{equation}
    P = \frac{T_p}{T_P + F_p} 
\end{equation}
\begin{equation}
    R = \frac{T_p}{T_P + F_n}
\end{equation}
where $T_p$ is the number of true positive craters, $F_p$ are false positives and $F_n$ are the false negatives. High precision implies that most craters \textcolor{black}{that} were found by the model are actually present in the ground truth, while high recall implies that the model was able to find most craters in the ground truth list (independent of false positive rate). 

In table~\ref{tab:results} we compare the precision and recall to DeepMoon\footnote{Note that DeepMoon's test set extended from +60 to +180 degrees longitude, while in this work the test set is between -60 and -180 degrees longitude.}. We follow DeepMoon in calculating two sets of recall/precision: Post-CNN where the values are \textcolor{black}{separately} calculated for \textcolor{black}{each of} the individual poststamp images \textcolor{black}{before being} averaged over the entire test set; and Post-processing where they are calculated \textcolor{black}{from} the \textcolor{black}{global {unique}} crater distribution. Post-CNN recall allows us to quantify the performance of the model on individual small scale images (how many known craters per image MaskRCNN is finding), while Post-processing recall gives us the overall performance of the model on the entire test set (how many of all known test set craters it is finding). Note that by transforming from a set of individual images to a global catalog we are essentially doing significant ensembling, giving the same crater multiple chances of being detected in at least one of the images.

\newcolumntype{P}[1]{>{\centering\arraybackslash}p{#1}}
\begin{table}
	\small
	\begin{center}
		\begin{tabular}{ |P{3.2cm}|P{2.1cm}|P{2.1cm}|P{2.1cm}|P{2.1cm}| } 
			\hline 
			Accuracy Metric & MaskRCNN (Post-CNN) & MaskRCNN (Post-proc) & DeepMoon (Post-CNN)& DeepMoon (Post-proc) \\ \hline
			Recall & 87.6 $\pm$ 8\% & 85.1\% & 57 $\pm$ 20\% & 92\% \\
			Precision & 66.5 $\pm$ 17\% & 40.2\% & 80 $\pm$ 15\% & 56\%  \\
			F1 &  0.75 & 0.54 & 0.66 & 0.69\\
			Frac. long. error &  10.5\% & - & 10\% & -\\
			Frac. lat. error &  7.5\% & - & 10\% & -\\
			Frac. rad. error &  7\% & - & 8\% & - \\
			\hline
		\end{tabular}
		\caption{Accuracy metrics table comparing the test set precision, recall, F1 score, and fractional error on craters coordinates of MaskRCNN to DeepMoon. Post-CNN is the average of the values calculated for each poststamp image, and is hence biased by craters present in more than one image. Post-processing is the value calculated over the entire test set once unique craters have been extracted, and their physical dimensions retrieved. }
		\label{tab:results}
	\end{center}
\end{table}

For post-CNN, we find that MaskRCNN has a 20\% higher recall than DeepMoon (87.6\% compared to 54\%), with a 12\% lower variance. This implies that MaskRCNN is significantly more robust in detecting craters in any given image, successfully identifying on average 87.6\% of all known craters. After post-processing however, MaskRCNN has found in total 85.1\% of all craters in the test set, while DeepMoon has found 92\%. {While we should note that the two models used different longitudes to define their test sets}, Fig. \ref{fig:size} shows clearly that DeepMoon is better at detecting intermediate sized \textcolor{black}{($\sim$15-40~km)} craters than MaskRCNN, probably leading to the better overall recall.  % \textcolor{blue}{[APJ: I wonder if there is also any influence from under-sampling at the smallest scales.  There is a very clear roll over in the size distribution at around 6 km that is indicative of incompleteness.]}
{It is interesting to note that this size range is roughly centred on the 20~km range where, as we discussed in Section~\ref{sec:data}, there is a known deficit in the P+H database.  DeepMoon does not appear to have been influenced by this deficit, whereas MaskRCNN may have been, suggesting that MaskRCNN may be more sensitive to quirks in the training data. Moreover, this size range coincides roughly with the simple to complex crater morphology transition, and hence an alternative explanation is that DeepMoon is better than MaskRCNN in detecting complex craters.}

The post-processing precision of MaskRCNN is 40.2\%, compared to 56\% for DeepMoon. However, as \cite{alidib} argued, the low precision score for these models is due to the machines identifying a large number of `new' craters {(i.e. craters not present in the GT test set)}.  \textcolor{black}{This is not surprising since the crater identification methodology of \citet{povilaitis2017} was conservative, including only surface features definitively identified as craters.} MaskRCNN's lower precision can hence be interpreted as the machine identifying more new craters with respect to the number of ground truth craters than DeepMoon. {We emphasize that we are comparing the performance of the two models on different test sets generated from different longitude ranges, and so one should be {cautious} when interpreting these results.}

As an ultimate test for the performance of our model, we calculated the false positive rate of ``new'' craters identified but not present in the ground truth catalogues. This was done in the same fashion as \cite{alidib} where a group of 3 astronomers were shown a representative random sample of data (30 poststamp images containing a total of 349 uncatalogued craters), and tasked with counting what they subjectively deemed as false detections. We then calculated the arithmetic mean of the 3 estimates to finally get a false positive rate of 24\%. 

\textcolor{black}{We emphasise that the `new' craters identified by the machine are not necessarily (and in most cases are unlikely to be) new to science and may well have been examined by previous studies, especially regional ones.  The key point here is that they are not present in the ground truth P+H data set and thus represent the machine applying the patterns it has learned from the craters contained in the ground truth and identifying additional surface features that satisfy them.}

{While global metrics are useful, it is important to characterize the performance of the CNN for different crater size bins. We hence calculate the diameter-dependent recall scores, and summarize them in table \ref{tab:2}. We find that the recall is consistently high ($\geq 90\%$) for craters between 10-30 km in diameter, then decreases to 82\% for smaller craters (due to the intrinsic difficulty in identifying small objects). The recall also decreases gradually for craters larger than 30 km, as these are less common in the GT, but also because their pixel sizes are larger than the network's convolutional filters.}

%\newcolumntype{P}[1]{>{\centering\arraybackslash}p{#1}}
\begin{table}
	\small
	\begin{center}
		\begin{tabular}{ |P{3.2cm}|P{1.5cm}|P{1.1cm}|P{1.3cm}|P{1.1cm}|P{1.5cm}|P{1.5cm}| } 
			\hline 
			Accuracy Metric & 5-10 km & 10-15 & 15-20 & 20-30 & 30-100 & 100-500 \\ \hline
			Recall (Post-CNN) & 82.5\% & 90.5\% & 90.5 \% & 90.0\%  &83.5\% & 72.2\% \\
			Recall (Post-proc) & 82.2\% & 90.1\% & 90.1 \% &89.4\%  &82.8\% & 64.4\% \\
			\hline
		\end{tabular}
		\caption{{The test set recall scores of MaskCNN for different diameter bins.}}
		\label{tab:2}
	\end{center}
\end{table}

{We note that the intermediate sized, 10-30~km craters are likely to be detectable at all of the levels of downsampling used in our poststamp generation pipeline, whereas the smaller 5-10~km craters will be harder to detect at the largest scales and the larger craters will be greater than the poststamp size (and thus undetectable) at the smallest scales.  This effectively means that the intermediate-sized craters have more chances to be detected and so the slightly higher recall here is probably not surprising.}

{Craters larger than 100 km in diameter have a significantly older average age, and therefore have undergone more erosion and resurfacing. As discussed previously in Section~\ref{sec:postproc}, our model is also insensitive to craters significantly larger than 100~km and so we should expect it to perform significantly more poorly in this size range. Since the number of craters in this size range is small we inspect the performance of our model on these craters by manually inspecting all of the craters larger than 100~km in diameter (68 unique ones in total in the test set). We find 13 false positives (many of which are multi-ringed basins that confused the CNN), and 5 craters not present in the GT. The false detection rate for large craters is then $\sim 20\%$. We finally calculate the recall for this size bin and find 72\% for post CNN and 64\% for the unique distribution.  }

Finally, alongside the recall and precision that tell us how good the machine is at identifying the presence of a crater, we also wish to examine how well the machine reproduces the dimensions and location of the crater.  Averaging over all detected craters that can be identified with a target crater in the ground truth we find that the fractional error in the centroid location of the crater has a standard deviation of 10.5\% of the crater radius in longitude and 7.5\% of the crater radius in latitude.  The difference between the longitude and latitude errors is likely due to the effects of the projection. For the radius we find that the standard deviation over the whole dataset is 7\%. {These values are consistent with the uncertainties calculated through comparing results from different human experts \citep{robbins2014}.}

To test for any size or location dependent biases we show in Table~\ref{tab:dRvslat} the fractional error in the radius as a function of latitude and in Figure~\ref{fig:Rrad} the radius as predicted by the machine versus the radius in the ground truth.  {Table~\ref{tab:dRvslat}} shows that there is no correlation between the fractional error in the radius and the latitude, reassuring us that in the $\pm60^{\circ}$ latitude range we examine there are no residual effects from the projection correction.  \textcolor{black}{Our model, however, does predict slightly more craters in the northern latitudes.  This is consistent with the ground truth where the northern hemisphere contains around 9\% more craters than the south, but we note that other catalogs find a larger number of craters in the southern hemisphere.  This difference may be due to a preference (in both the ground truth and the machine predicted databases) for more clear-cut (less degraded) crater candidates.} From Figure~\ref{fig:Rrad} we can see that our model does have a slight tendency to underpredict the crater radius for craters larger than around 20~km, and that the dispersion is somewhat larger for these larger craters.

{Note that while crater overlap can contribute to these errors if treated as an individual crater, it should not preferentially result in errors in either latitude or longitude.  Unidentified overlap will contribute to the mean error in centroid location, but this is infrequent enough that the contribution is minor.}

%Finally we quantify the accuracy of the identified craters and test for any size or location-dependant detection bias by plotting the predicted radii vs the ground truth radii of our post-processed craters distribution, in addition to the fractional error on the radius as a function of ground truth latitudes, in Figures and \ref{fig:Rrad} and \ref{fig:dR} respectively. The predicted vs GT radii plot shows that our model has a slight tendency to underpredict the radius for values larger than 20 km, but this should have no effects on our results. The fractional error on radius we plot is uncorrelated with the latitude, showing that there is no residual effects left from the projection correction.

%\begin{figure}
%	\begin{center}
	%	\centerline{\includegraphics[scale=0.3]{dRvsLat.png}}
	%	\caption{The fractional radius error between the ground truth and predictions defined as $abs(R_{G}-R_{P})/\Bar{R}$, as a function of the GT craters latitudes. Colors are for display purposes only. The uniformity of the scatter shows that the detection precision is {to a large degree} uncorrelated with latitude. {Our model however does predict slightly more craters in the northern latitudes, which is consistent with the ground truth where the northern hemisphere contains $\sim$ 9\% more craters. }}
	%	\label{fig:dR}
%	\end{center}
%\end{figure}

%\newcolumntype{P}[1]{>{\centering\arraybackslash}p{#1}}
\begin{table}
	\small
	\begin{center}
		\begin{tabular}{ |c|c|c|c|c|c|c| } 
			\hline 
			Latitude range & -60 to -40 & -40 to -20 & -20 to 0&0 to 20 & 20 to 40 & 40 to 60 \\ 
			frac. dR$\leq$ 0.04 & 90.2\% & 87.7\% & 90.2 \% &89.1\%  &87.1\% & 89.3\% \\
			\hline
		\end{tabular}
		\caption{{Fraction of craters where the fractional radius error between the ground truth and predictions defined as $abs(R_{G}-R_{P})/\Bar{R}$ is less than 0.04, for different GT craters latitude ranges. As the lowest value found is 87.1\%, we conclude that the detection precision is {to a large degree} uncorrelated with latitude. %Our model, however, does predict slightly more craters in the northern latitudes, which is consistent with the ground truth {where the northern hemisphere contains $\sim$ 9\% more craters (though note that other catalogs show a larger number of craters in the southern hemisphere). A possible alternative explanation is that MaskRCNN performs better on well preserved craters, in contrast with the more degraded craters that dominate the highlands.}
		}}
        \label{tab:dRvslat}
	\end{center}
\end{table}

{As we identified in Section~\ref{sec:data}, while we do not use the database of \citet{robbins2019} for training our model, it can still be useful as an additional source of testing and comparison.  In table~\ref{tab:robbinscomp} we compute the `recall' for the post-processed output of MaskRCNN compared to the \citet{robbins2019} database, using the same radius bins as for the recall calculation with the ground truth P+H database.  In addition, we make use of the arc fraction parameter provided by \citet{robbins2019}, cutting the database at three different values of the arc fraction and computing the recall against those slices.  The arc fraction is the fraction of a complete circle or ellipse used when fitting the crater.  A fresh, young crater would be expected to have an arc fraction of 1.0 or very close to that value, while older craters that have been eroded and overprinted by other craters and lost parts of their crater rims will have lower arc fractions.  The arc fraction can thus be taken as a loose proxy of how degraded the crater is and thus a rough guide to how confident we can be about the fit and the identification of the feature as a crater.}

\begin{table}
    \centering
    \small
    \begin{tabular}{|c|c|c|c|c|c|c|}
    \hline
         Arc fraction & 5-10~km & 10-15~km & 15-20~km & 20-30~km & 30-100~km & $>$100~km \\
         \hline
         $>$0.95 & 81.3\% & 91.9\% & 95.1\% & 92.8\% & 87.2\% & 100\%$^+$ \\
         $>$0.75 & 48.1\% & 81.6\% & 86.3\% & 88.6\% & 88.0\% & 75.0\% \\
         $>$0.5  & 35.6\% & 65.2\% & 73.7\% & 80.6\% & 84.3\% & 70.3\% \\
         \hline
    \end{tabular}
    \caption{{Recall (for post-processed output) for MaskRCNN compared to the \citet{robbins2019} database with three different cuts based on the arc fraction parameter. $^+$There were only two craters in this bin.}}
    \label{tab:robbinscomp}
\end{table}

{What we can see from table~\ref{tab:robbinscomp} is that for craters in the \citet{robbins2019} database with the most complete rims the recall is very similar to, or marginally better than, the recall we determine for the P+H database on which the machine was trained.  Moving to craters with up to a quarter of the rim missing we can see that the recall is still comparable to that obtained for the P+H database for craters larger than 10~km in diameter.  The exception is craters in the 5-10~km range where the recall has dropped dramatically.  We should not be surprised by this large drop however when we remember the differences in the raw data from which the databases are derived.  The \citet{robbins2019} database was compiled using optical imagery of significantly higher resolution than the DEM on which we are training the machine, this is how they were able to identify craters as small as 1~km.  Moving from a higher resolution to a lower resolution the first craters that we would expect to pass from detectable to undetectable are the smallest, most degraded ones as in the lower resolution imagery the subtle features that indicate that an incomplete feature is indeed a crater are harder to discern.  We see the same behaviour when we further include all craters in the \citet{robbins2019} database with an arc fraction of $>$0.5, with the recall on 5-10~km craters continuing to fall and the 10-15~km range now beginning to drop substantially while the other size bins fall by lesser amounts.}

{The behaviour of the \citet{robbins2019} recall as a function of size and arc fraction reinforces our discussion in Section~\ref{sec:data} regarding our reasoning for using the P+H database for model training.  Since smaller craters are larger in number using the \citet{robbins2019} database would likely have resulted in a very large number of craters that the machine would have had great difficulty identifying from the DEM input which would likely have made training much more difficult.  The recall as a function of size and arc fraction is also interesting as it gives us information on how the degradation state of a crater influences the ability of the machine to detect it relative to the resolution of the input data. {Finally, we now calculate the number of craters that were identified by the machine but were not present in the catalogues, and compare it between the two different ground truths. For the P+H dataset, $\sim 10000$ such ``new'' craters were identified, compared to 7000 when considering the \cite{robbins2019} catalogue with arc fraction larger than 0.5. Note that for lower arc fractions, the matching crater numbers increases much slower, then plateaus. We interpret this as the hyper-parameters of the recall assessment module (equivalent to solving equation \ref{eq1} for the GT and predicted distributions) being fine tuned for DEMs, not optical imagery.  }}

\subsection{Crater size distribution}
\label{sec:diam}
\begin{figure}
	\begin{center}
		\centerline{\includegraphics[scale=0.4]{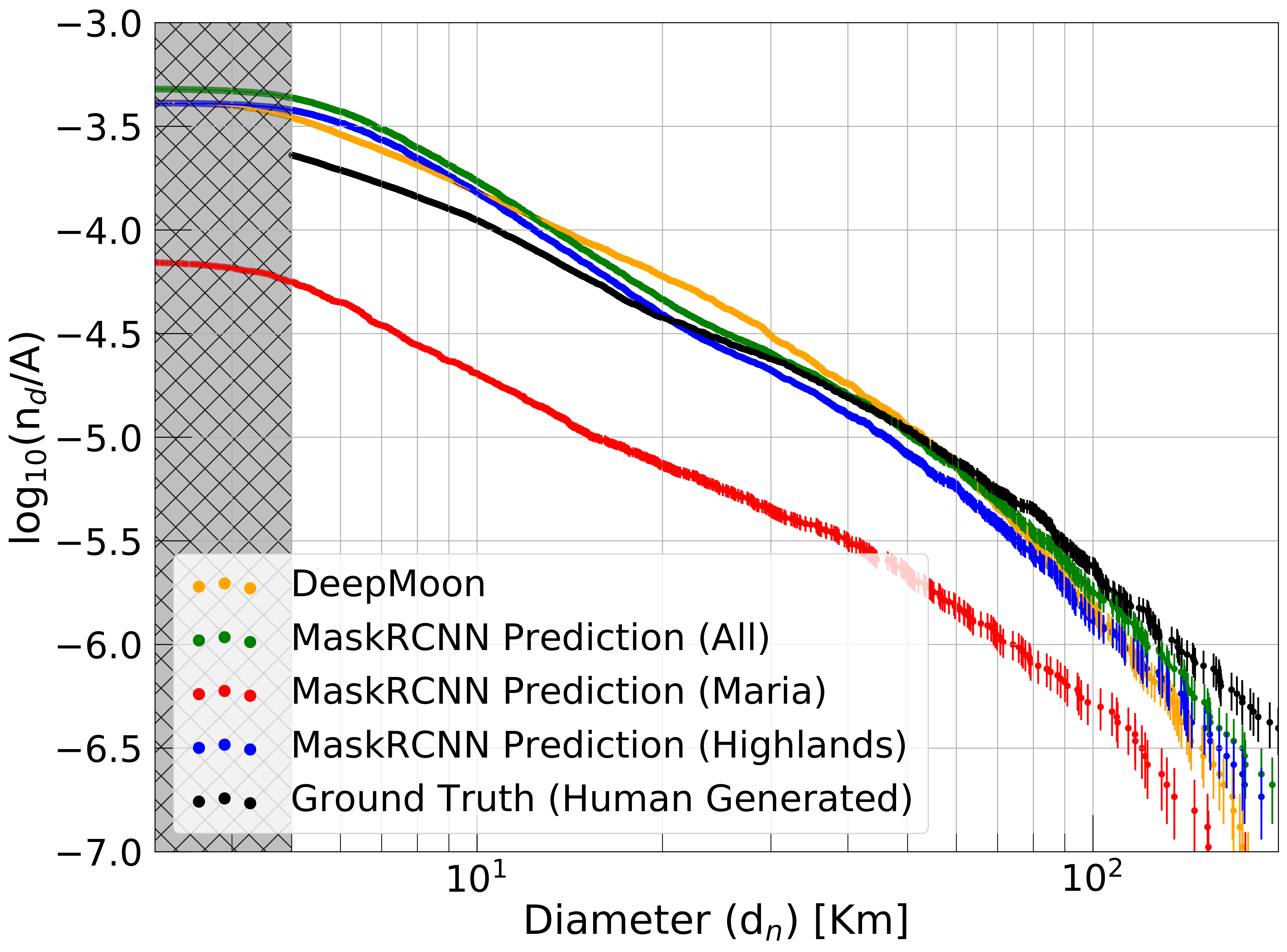}}
		\caption{Lunar crater size-frequency distributions represented as CSFD plots, for our test set. Black is the human-generated ground truth, red and blue are MaskRCNN's predictions for the maria and highlands craters, respectively, and green is both combined. The model predictions follow the GT closely for intermediate and large craters, before increasing more steeply for small craters where MaskRCNN is detecting new craters not present in the GT. {When comparing DeepMoon and MaskRCNN it should be emphasized that their test sets used different longitudes.} }
		\label{fig:size}
	\end{center}
\end{figure}

\begin{figure}
	\begin{center}
		\centerline{\includegraphics[scale=0.3]{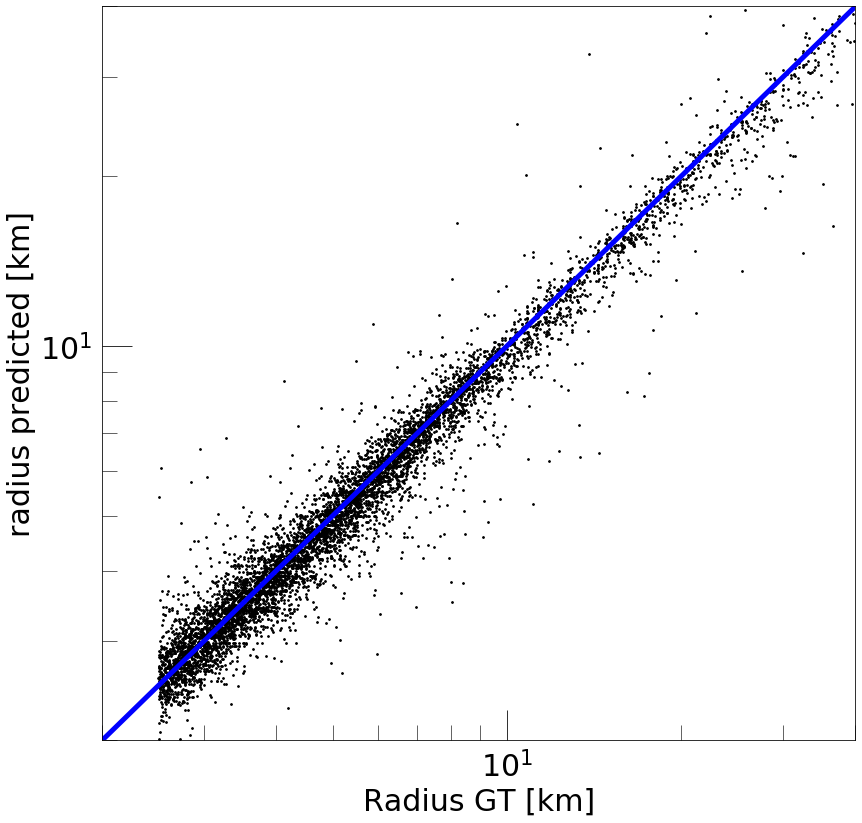}}
		\caption{The predicted craters radii as a function of the GT radii. Scatter points distribution is symmetric with respect to the blue diagonal for craters smaller than $\sim 20$~km, implying that the fractional radius error is independent of the craters size in this range. }
		\label{fig:Rrad}
	\end{center}
\end{figure}

The major success of DeepMoon was the accurate retrieval of the test set's crater size-frequency distribution. In this section we analyze the performance of MaskRCNN on this fundamental task. In Fig. \ref{fig:size} we show the crater size distribution predicted by MaskRCNN presented as a cumulative size-frequency distribution (NASA Technical Memorandum number 79,730 (1978)). We show the separate distributions of the lunar highlands and maria, compared to the ground truth from our test set. The predicted distribution closely follows the GT for craters larger than 30 km, while increasing more steeply for smaller diameters. This is due to MaskRCNN ``discovering'' a large number of smaller craters not present in the GT. One of the shortcomings of DeepMoon was the decrease in detection performance for craters larger than 40 km. 
\textcolor{black}{Comparing MaskRCNN and DeepMoon curves in Fig.~\ref{fig:size} we can see that although there is a slight drop-off in the performance of MaskRCNN at larger crater diameters it is a marked improvement over DeepMoon for these sizes}.  MaskRCNN however detects less intermediate sized craters with diameters between 15 and 40 km than DeepMoon (even though it is still finding more craters than in the ground truth), while detecting more for diameters smaller than 15 km. 

This highlights the importance of models ensembling in machine learning, since different models can have better performance in different regimes. Note that the distributions inferred by both DeepMoon and MaskRCNN {plateau} for craters smaller than 5 km, which is a sign of {GT incompleteness and the intrinsic difficulty in identifying craters} at these scales. {We can also see the distribution of highlands craters begin to roll over at around 7~km, whereas the mare distribution only begins to roll over below 5~km.  This difference is likely because the lower density of craters on the mare results in less confusion and easier detection at these small scales.  The highlands distribution (and also the total distribution) beginning to roll over above 5~km is consistent with table~\ref{tab:2}, which shows that recall begins to drop slightly below 10~km.}

In Fig. \ref{fig:size} we can also infer that the cumulative size distributions of highland and maria craters are \textcolor{black}{very similar} except at very large sizes where \textcolor{black}{the maria distribution suffers from low number statistics}, and very small sizes where \textcolor{black}{the distributions plateau due to resolution limitations}. This is demonstrated further with a {2-sample Kolmogorov-Smirnov test for craters larger than 10 km (where the GT can be assumed to be reasonably complete), returning a p-value of 0.17 {and thus the null hypothesis that the two populations are the same cannot be rejected .}}     

\subsection{Lunar Crater Ellipticity Distribution}
\label{sec:csfd}

\textcolor{black}{Not all craters are exactly circular, impacts at a non-normal angle can result in elliptical craters, elongated along the direction of travel of the impactor \citep[e.g.][]{gault}.  The expected distribution of impact angles is well known ($P(\theta)d\theta = 2\sin 2\theta d\theta$, \citealt{shoemaker1962}), and peaks at an angle of 45$^{\circ}$, diminishing towards both perfectly vertical and perfectly grazing.  Given a large set of impact craters however it is expected that a fraction of them will occur at shallow enough angles to generate significant ellipticity. The well defined distribution of expected impact angles allows the ellipticity distribution to be mapped onto the angle distribution to find the impact angle that results in a certain ellipticity.  The threshold angle at which a given ellipticity is achieved varies between different surfaces and depends on the strength of the target material \citep[e.g.][]{collins}.  Stronger materials more readily retain information about impact orientation and have a larger threshold angle for a given ellipticity.  The ellipticity distribution on the surface of a body is thus of interest since it provides some access to information about the strength of the surface materials.}

%In this section we analyze the crater shape distribution found by MaskRCNN, and compare it to previous works.

\begin{figure}
	\begin{center}
		\centerline{\includegraphics[scale=0.3]{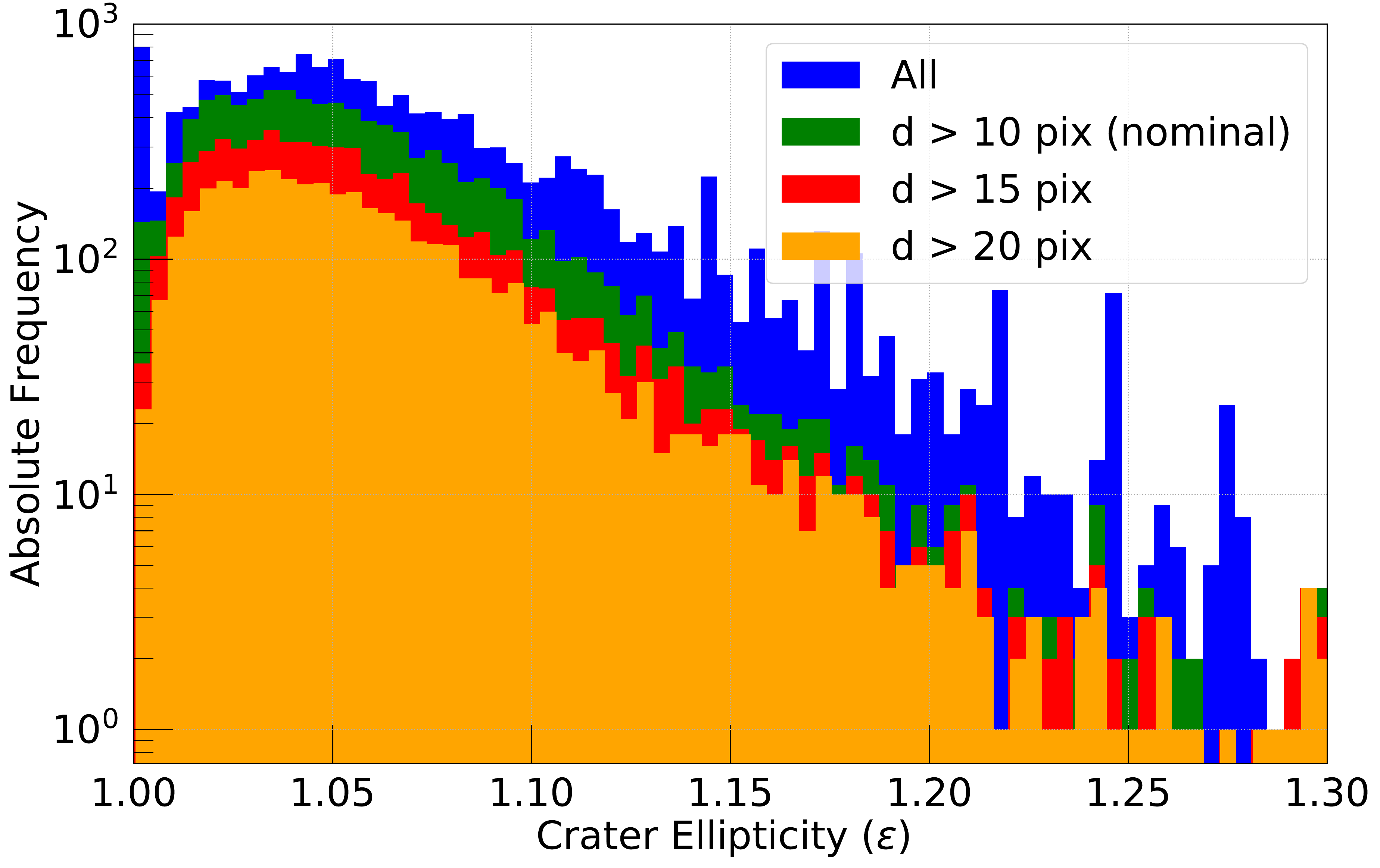}}
		\caption{Crater ellipticity frequency distribution as predicted by our model for the test set, for different levels of minimum crater diameters in pixels. Notice the spikes when all craters are considered, due to inaccurate ellipse fitting of the smallest craters resolved with a small number of pixels.  
		These disappear when we exclude craters smaller than 10 pixel in diameters.}
		\label{fig:ellippix}
	\end{center}
\end{figure}

\textcolor{black}{Ellipse fitting is easier with the mask approach that we use here compared with the binary rings used by \citet{alidib} and so here we examine the crater shape distribution found by MaskRCNN.  In Fig.~\ref{fig:ellippix} we show the ellipticity ($\epsilon$) frequency distribution of craters in the test set.  An issue for ellipse fitting with small craters is the quantisation of the crater dimensions into the image pixels.  For example, a crater that is 7 pixels along the longest dimension can be 7 by 7 ($\epsilon = 1$), 6 by 7 ($\epsilon = 1.16$), etc, but not values in between.  To deal with this issue we introduce a cut-off in the on-image uncalibrated diameter (in pixels) below which we consider a crater to be too small for accurate ellipse fitting and exclude it from the distribution.  The blue histogram in Fig.~\ref{fig:ellippix} shows all of the craters with no cut-off and has distinct spikes at $\epsilon = 1$ and multiple values {at and larger} than $\epsilon = 1.12$.  The green, red and orange histograms then show the distribution with successively larger cut-off sizes of 10, 15 and 20 pixels.  We can see that even the 10 pixel cut-off is sufficient to eliminate almost all of the spikes and the shape of the distribution is very similar for all three values.  As such for the rest of this section we use a cut-off size of 10 pixels.}

\textcolor{black}{Compared to standard crater counts {(that either ignore ellipticity or use an average diameter)} studies of elliptical craters are somewhat sparser.  The earliest systematic observational studies specifically of elliptical craters focussed on Mars \citep[e.g.][]{barlow1988}, while early lunar studies examined unusual or anomalous craters more generally rather than specifically elliptical craters \citep[e.g.][]{schultz1982}.  Surprisingly, the first study that explicitly examines elliptical lunar craters is probably that of \citet{bottke}.}
%\textcolor{blue}{\sout{Surprisingly, one of the earliest works quantifying the proportion of elliptical lunar craters was done relatively recently by \cite{bottke} (probably testifying to the complexity of the task).}}
In that work, the authors surveyed the lunar maria for elliptical craters using Lunar Orbiter IV images. They measured a total of 932 craters between 2.3 and 89 km in diameter, and concluded that \textcolor{black}{50 of those ($\sim 5.4\%$)} have an ellipticity higher than 1.2.
{This compares well with our data for which we find that $\sim 3\%$ (494 out of 16,664 craters) have $\epsilon > 1.2$.} 

{It is however important to assess the reliability of these numbers. We hence manually inspect 50 unique craters with $\epsilon > 1.2$, and find that $\sim$ 12\% are misidentifications due to tight overlap of multiple small craters. Note that this is not necessarily a systematic problem since, as can be seen in Fig. \ref{fig:detection}, MaskRCNN correctly identifies the masks of many overlapping craters. {While this issue is present for most crater sizes, it is more prominent for small (thus also shallow) craters.} This false detection rate can be hence improved by using bigger images at higher resolution, necessitating significantly more computing power.}

%\textcolor{blue}{\sout{More recently, \cite{collins} used numerical simulations based on lab measurements to predict that \textcolor{black}{2-4\%} of 5-100 km sized craters on the Moon should have an ellipticity higher than 1.1. Both works see weak (or no) correlation between the ellipticity and diameter for these sizes. In Fig. \ref{fig:ellepticity2} we show the ellipticity vs diameter distribution for the lunar highlands and maria separately. For lunar $\epsilon > 1.2$, our model predicts that 1.84\% and 2.51\% of the highlands and maria craters respectively are elliptical. \textcolor{blue}{[Give the value for $\epsilon > 1.1$ as well]} Taking into account the systematic uncertainties of the model discussed above places the fraction of elliptical craters found by MaskRCNN between 1.41\textcolor{black}{\%} to 3.30\textcolor{black}{\%} for highlands, and 2.06\textcolor{black}{\%} to 3.66\textcolor{black}{\%} for the maria, consistent with the observations and theoretical predictions.}}

\subsubsection{Ellipticity -- angle distribution}
\label{sec:ellangledist}

\begin{figure}
    \centering
    \includegraphics[width=\columnwidth]{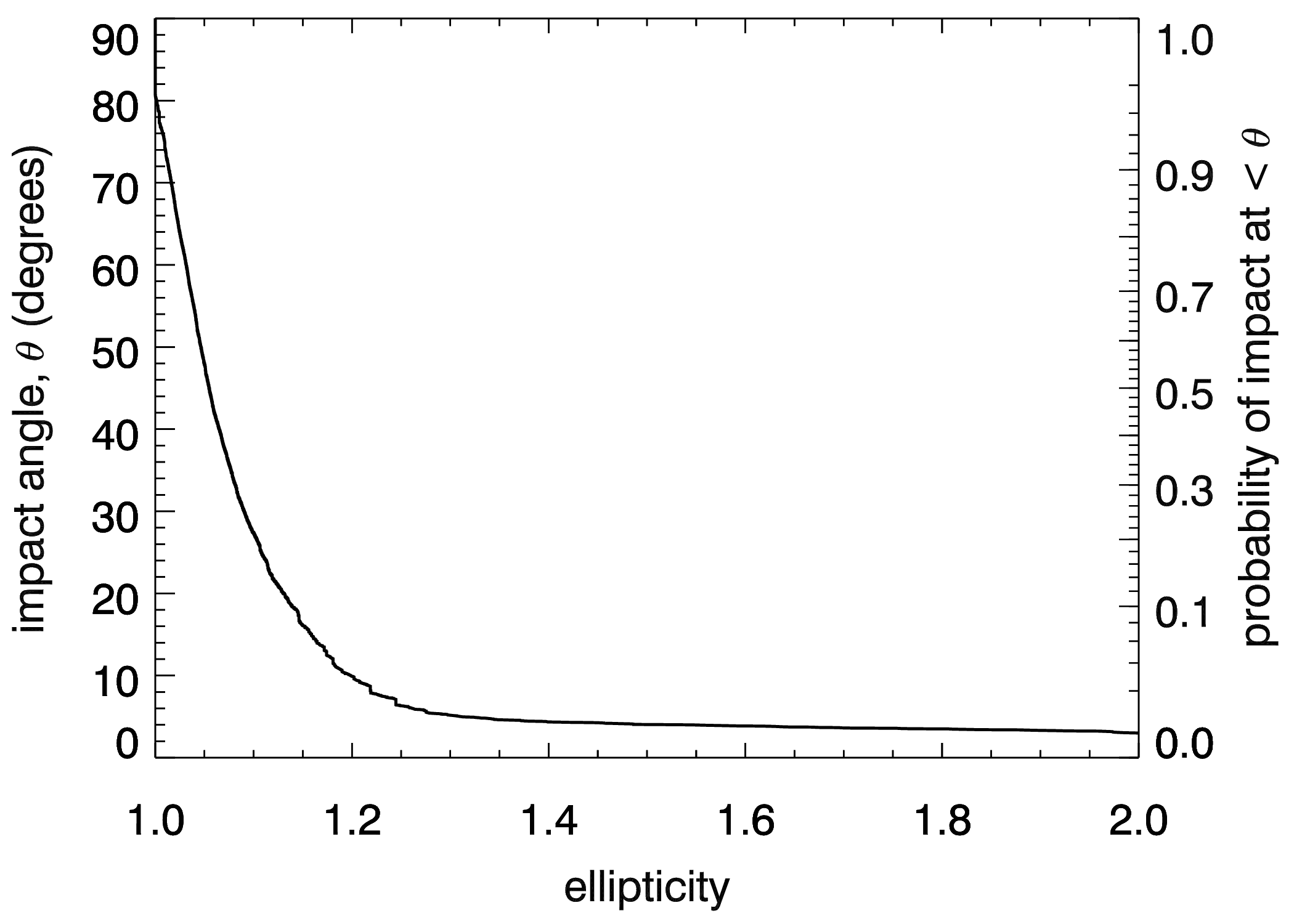}
    \caption{Ellipticity-impact angle distribution for the 16,664 craters in the test set, constructed assuming that the impact angle follows a standard $P(\theta)d\theta = 2\sin 2\theta d\theta$ distribution.  The left-hand axis shows the impact angle, $\theta$, and the right-hand axis shows the probability of an impact at $\leq\theta$.}
    \label{fig:ell-angle}
\end{figure}

Previous studies like that of \cite{bottke} have focussed on the proportion of craters that are more elliptical than a certain threshold, typically $\epsilon =$ 1.1 or 1.2.  This is partly because conducting a search for elliptical craters, fitting ellipses in a manual or only semi-automated fashion, is much more time-consuming than simple circular searches, especially for craters that are fairly close to circular.  {One can reasonably fit a circle by taking three points around the rim for example, whereas accurately fitting an ellipse requires {a minimum of five points}.}  Concentrating on craters that are unambiguously elliptical dramatically reduces the number of craters that the human counter needs to examine.  Our fully automated approach however allows us to examine many more craters, including the full distribution from circular to highly elliptical as illustrated in Fig.~\ref{fig:ellippix}.

In addition to converting the fraction with an ellipticity above a certain value into a threshold angle, having a large sample of ellipticities allows us to construct a complete ellipticity-angle distribution as shown in Fig.~\ref{fig:ell-angle}.  From this we can see that impacts at less than 27$^{\circ}$ result in $\epsilon > 1.1$, while impacts at less than 10$^{\circ}$ result in $\epsilon > 1.2$.  There are around 6 times as many craters with ellipticities in the range 1.1-1.2 than there are with $\epsilon > 1.2$.  This 27$^{\circ}$ threshold angle for $\epsilon > 1.1$ compares well with the results of \citet{collins} for numerical simulations of intermediate and gravity regime impacts with fairly low cohesion.  Note that while the expected impact angle distribution for primary craters is well known the distribution for secondary craters may differ from this.  However, we make no attempt to separate secondary craters and assume that all craters obey the expected impact angle distribution for primary craters in constructing Fig.~\ref{fig:ell-angle}. 

\textcolor{black}{Note that there are a small number (less than 0.3\%) of craters in our dataset that have ellipticities greater than 2, which may be a result of failures in the ellipse fitting, however, excluding these craters does not significantly change the distribution at intermediate ellipticities.  Similarly, around 3.5\% of the craters have ellipticities of exactly 1.  This is due to the limitations of pixel-based imaging noted above, even for a crater that is 100 pixels in the long dimension the lowest non-circular ellipticity it can have is 1.01, and is why the curve in Fig.~\ref{fig:ell-angle} appears to intersect the y-axis at 80$^{\circ}$ rather than 90$^{\circ}$.  As with the small tail of very high ellipticities excluding these does not significantly change the distribution at intermediate ellipticities.}

\subsubsection{Comparison with \citet{robbins2019}}
\label{sec:ellrobbins}

{Recently, \citet{robbins2019} used their extensive new dataset to extract statistics on the ellipticity of Lunar craters.  They found that, for craters larger than 10~km in diameter, 21\% have $\epsilon > 1.2$, 8.2\% have $\epsilon > 1.3$, and 1.4\% have $\epsilon > 1.5$.  Limiting our sample to the same diameter range we find that 3.2\%, 2.0\%, and 1.4\% of craters have $\epsilon > 1.2, 1.3, 1.5$ respectively.  \citet{bottke} use a different size range and restrict themselves to the lunar maria (their smallest crater being 2.3~km in diameter), but their sample of 932 craters contains 50 (5.4\%) with $\epsilon > 1.2$, 27 (2.9\%) with $\epsilon > 1.3$ and 14 (1.5\%) $\epsilon > 1.5$.  All three studies are in agreement at high ellipticities, but \citet{robbins2019} diverges substantially from \citet{bottke} and us at moderate ellipticities.}

{\citet{robbins2019} themselves noted this discrepancy with \citet{bottke} and put forward a number of possible explanations.  Several of these relate to the differences between the samples used since \citet{bottke} had a small sample restricted to the lunar maria and selected against secondary craters, whereas \citet{robbins2019} is a large global study and agnostic to whether craters are primary or secondary.  However, our sample shares most of these differences and yet finds fractions of elliptical craters that are more similar to \citet{bottke}.  As such, we suggest that the discrepancy between \citet{robbins2019} and \citet{bottke}, and our study, is unlikely to be due to the lack of secondary or highlands craters in \citet{bottke}.}

\begin{figure}
    \centering
    \includegraphics[width=\columnwidth]{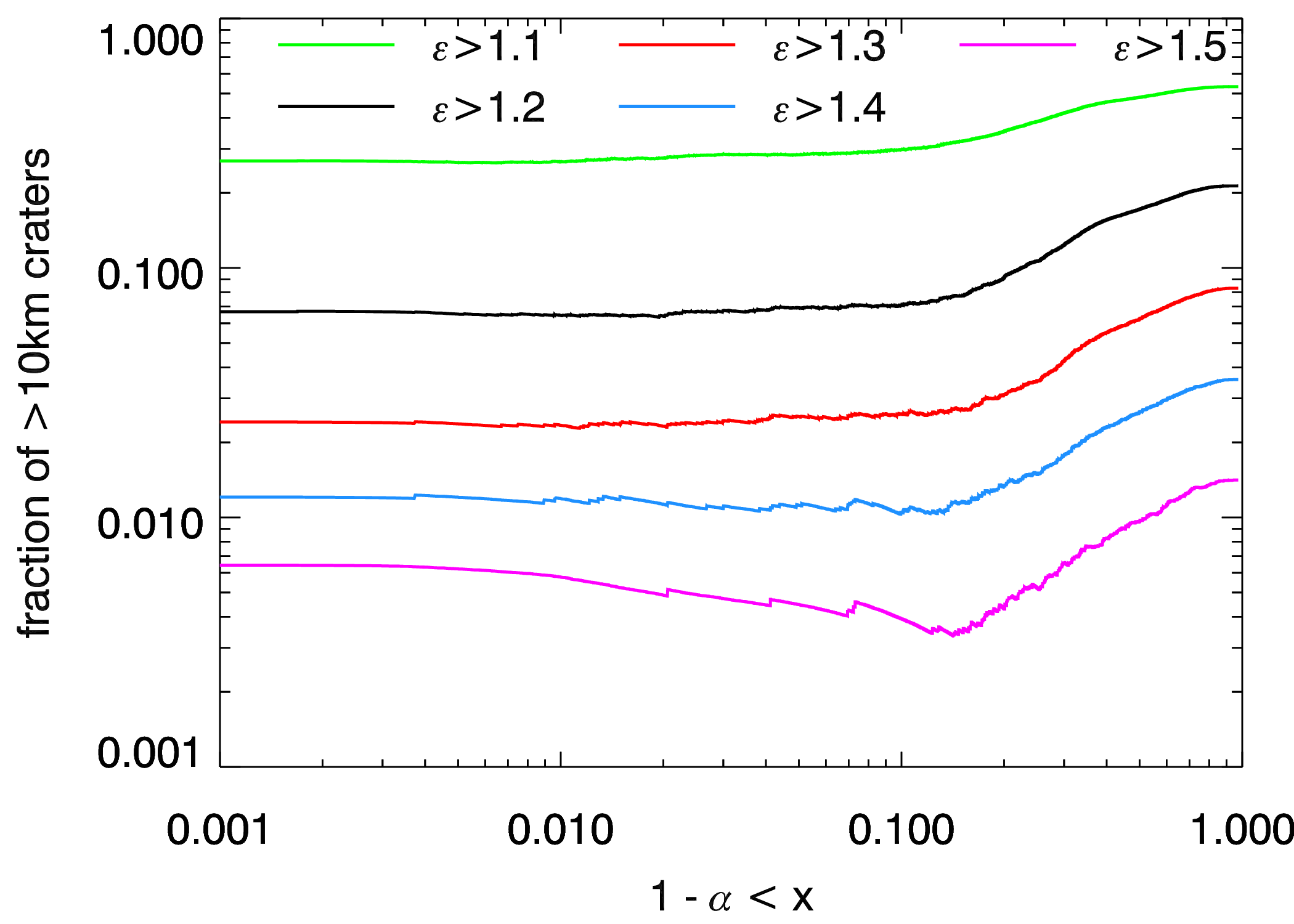}
    \caption{The fraction of craters in the \citet{robbins2019} database larger than 10~km in diameter that exceed a given ellipticity threshold as a function of the arc fraction, $\alpha$.  A point on the $x$-axis contains all craters for which $1 - \alpha < x$, i.e. the point at $x$=0.1 contains all craters which have $1 - \alpha < 0.1$, or identically all craters for which $\alpha > 0.9$}
    \label{fig:robbinsellfrac}
\end{figure}

{An important factor, noted by \citet{robbins2019}, may be that their ellipticities show a strong bias for fits to be more elliptical when a smaller segment of the crater rim is traced.  This is illustrated in figure~\ref{fig:robbinsellfrac} where we show how the fraction of craters that exceed 5 different ellipticity thresholds varies as a function of $(1-\alpha)$, where $\alpha$ is the arc fraction - the fraction of the full ellipse used to trace the crater rim.  As we can see there is a strong correlation between the fraction of craters more elliptical than a given threshold and the arc fraction for $\alpha \lesssim 0.85$ ($1-\alpha \gtrsim 0.15$).  Moreover for ellipticity thresholds of $\epsilon=1.2$ and greater the behaviour is the same with the lines lying roughly parallel to one another, in each case rising by a factor of around 3 from $1-\alpha = 0.1$ to $1-\alpha = 1.0$ as we include craters that have less and less rim arc used for fitting.  When greater than 90\% of the rim is available for fitting however, the fraction of craters more elliptical than a given threshold is quite stable (the slightly different behaviour of the $\epsilon > 1.5$ line at high arc fractions is due to the small numbers here).  If we take these stable values then we find that the \citet{robbins2019} database has 6.5\%, 2.4\% and 0.55\% of craters with $\epsilon > 1.2, 1.3$ and $1.5$ respectively.  These values are much closer to those in \citet{bottke}, and our study, albeit that they are now somewhat lower for the most elliptical craters.}

{While the similarity of the elliptical crater fractions when we only include craters with rims that are $>90\%$ complete hints that this may be the solution to the differences between \citet{robbins2019} and other studies it is not definitive.  We do not have equivalent information for \citet{bottke} so it is possible that the elliptical crater fraction would also fall for the \citet{bottke} sample if we excluded craters with rims that are less than $90\%$ complete.  Nonetheless, the lower crater density in the maria, and the exclusion of more eroded craters, does mean that the \citet{bottke} sample is likely to have contained a larger proportion of complete craters than \citet{robbins2019}.}

{Our algorithm works in a fundamentally different way than \citet{robbins2019} or \citet{bottke} so there is no direct comparison for the arc fraction parameter of \citet{robbins2019}.  We can however construct a crude analogue by dividing the craters in our sample into those which overlap and those which do not, since crater overlap is a major cause of segments of the rim being missing.  Given two craters with centres at $(x_1, y_1)$, $(x_2, y_2)$ and radii $r_1$, $r_2$, where $r_1>r_2$, their rims will overlap if}
\begin{equation}
    (r_1 - r_2)^2 < (x_1 - x_2)^2 + (y_1 - y_2)^2 < (r_1 +r_2)^2.
\end{equation}
{This criterion excludes craters for which the separation of the centres is smaller than the difference in the radii since these will be completely nested inside one another with no rim overlap.  Note that this equation does assume that the craters are circular, but for moderate ellipticities the difference is small.}

\begin{figure}
    \centering
    \includegraphics[width=\columnwidth]{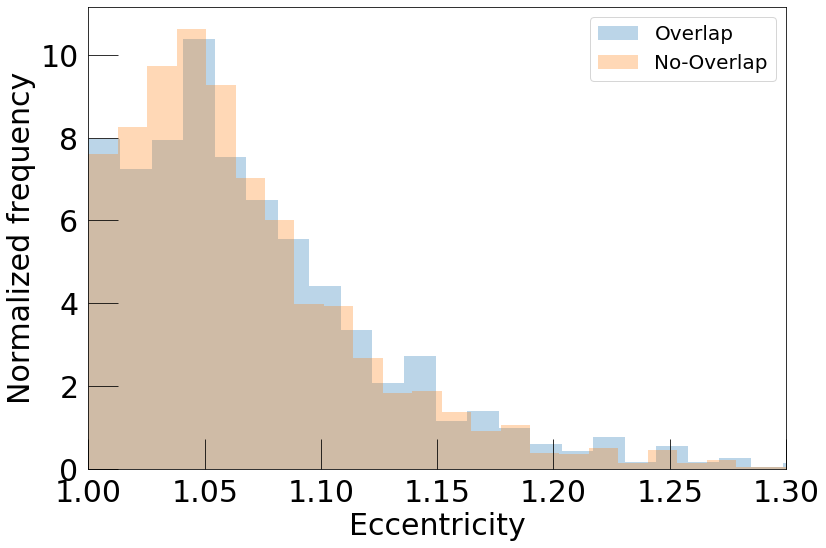}
    \caption{Histograms of crater ellipticity for craters identified by our algorithm, divided into craters that do not overlap (yellow) and those which do (blue).}
    \label{fig:rimoverlap}
\end{figure}

{In figure~\ref{fig:rimoverlap} we implement the split between craters that do, and do not, overlap in the sample of craters identified by our algorithm and show the distribution of ellipticities for the two sub-samples.  As we can see there is no discernable difference between the sub-samples.  While the analogy between the division based on overlap and the arc fraction is far from perfect (and overlap may exist in the non-overlapping set with craters not identified by our algorithm), this suggests that our ellipticities may be less influenced by the presence of overlapping craters and partial rims.  We speculate that this may be because to produce ellipticities we are fitting to the perimeters of the masks, and these are always complete with points roughly evenly spaced even if the rim of the underlying crater is not complete.}

{The machine is also less likely to recognise a crater for which the rim has very large portions missing due to overlap or erosion, as we identified with the recall comparison to \citet{robbins2019} in Section~\ref{sec:iden}.  This means that as with \citet{bottke} the typical crater in our sample is likely to be less eroded than one in \citet{robbins2019}.  Since the machine was originally trained on circular masks it would also not be entirely surprising if it tended to favour a more circular prediction when information about the rim was significantly incomplete.  Finally, it is important to note that, as identified by \citet{robbins2019}, the bias for higher typical ellipticities with decreasing arc fraction may be at least partly real, in that it is possible that a preferential erosion mechanism could make older, more eroded craters appear more elliptical on average than younger, fresher ones.}

{Biases in ellipticity fits in large crater databases is clearly an area of interest and one in which the interaction of machine generated databases like ours and human generated databases like that of \citet{robbins2019} may be fruitful.}

\subsubsection{Comparison of highlands and maria craters}
\label{sec:ellhighvsmaria}

\begin{figure}
	\begin{center}
		\centerline{\includegraphics[scale=0.3]{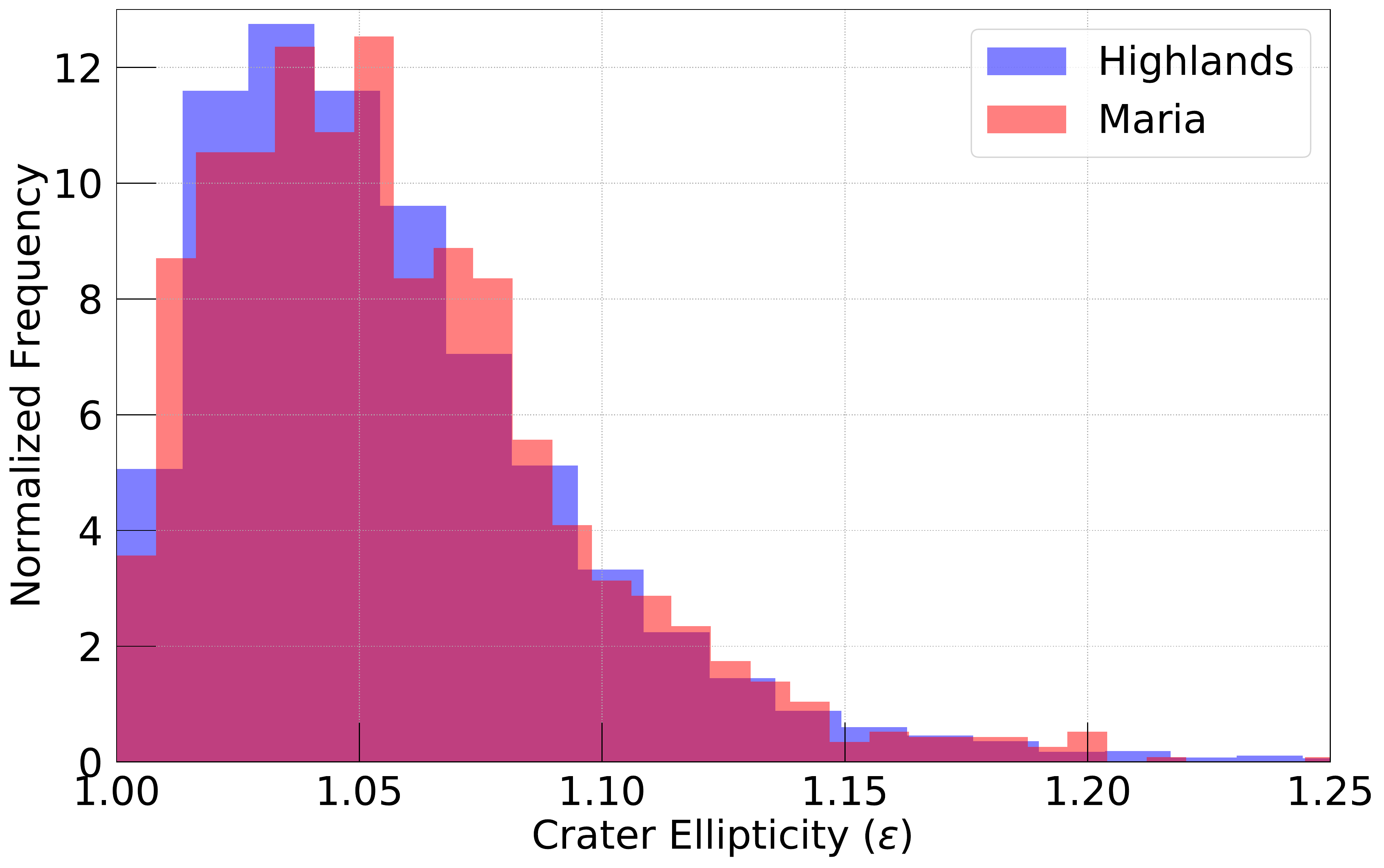}}
		\caption{Crater ellipticity frequency distribution for the highlands and maria, normalized for ease of comparison. This reveals no notable differences between the two. }
		\label{fig:ellepticity}
	\end{center}
\end{figure}

\textcolor{black}{As we have discussed, the relationship between impact angle and crater ellipticity depends on the strength of the target materials.  As such it is interesting to compare the ellipticity distributions for different regions of the Moon as differences may indicate differences in the properties of the surface materials.  The most obvious distinction in surfaces on the Moon is between the maria and the highlands, and so in Fig.~\ref{fig:ellepticity} we separate the crater populations of the maria and highlands and show the ellipticity frequency distributions for each.  We can immediately see that the distributions are very similar except perhaps at the high ellipticities where the maria suffer from low number statistics.  A statistical test confirms this, with a 2-sample Kolmogorov-Smirnov test returning $D=0.146$, implying that both samples were probably drawn from the same distribution.  This is not surprising since the highlands inevitably dominate any global crater distribution and we have already noted the compatibility between our total results and those of \citet{bottke} that were derived only from maria craters.}
%{Noah: To me, figure 6 shows that Highlands craters have a higher average elipticity than Mare craters. This is great since its consistent with predictions. Even if the difference is small, I think we should point it out.}
%[APJ: As the new figure 7 shows, the distributions are consistent at all sizes, the larger number of blue points tends to make it seem like the proportion of elliptical craters is larger because you can't see the individual points in the dense regions.]

\begin{figure}
	\begin{center}
		\centerline{\includegraphics[scale=0.3]{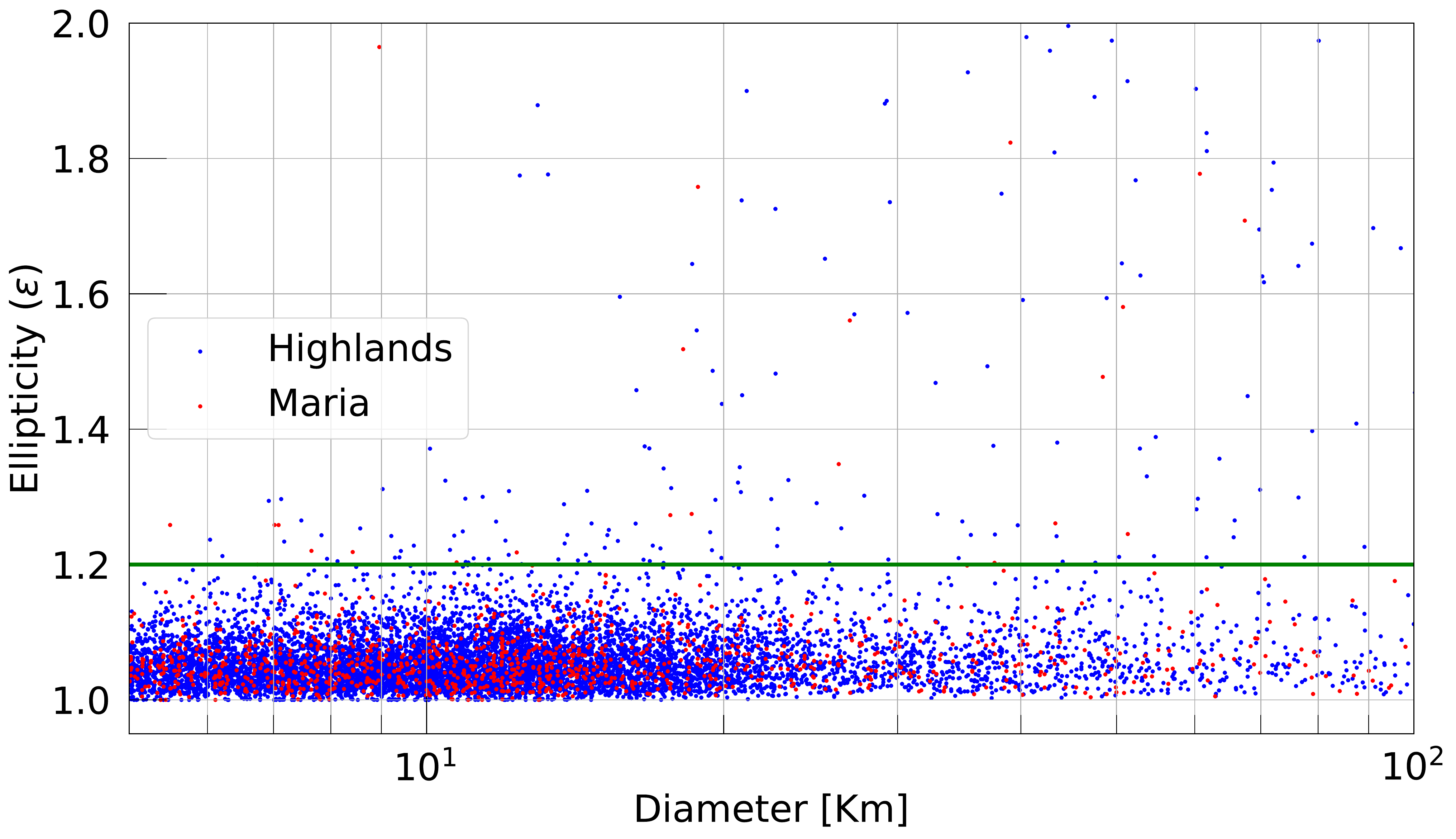}}
		\caption{Ellipticity (ratio of major to minor axis) of the test set craters as a function of diameter, as predicted by MaskRCNN. We cut off this plot at 5 km, to limit the effects of artifically distorted small craters (with diameters comparable to pixel size) on the distribution. Red dots are for maria, and blue for highlands. The ratio of elliptical (\textcolor{black}{$\epsilon > 1.2$}) to non-elliptical craters predicted is consistent with the observational results of \cite{bottke}. }
		\label{fig:ellepticity2}
	\end{center}
\end{figure}

\subsubsection{Ellipticity as a function of crater size}
\label{sec:ellsize}

\begin{figure}
    \centering
    \includegraphics[width=\columnwidth]{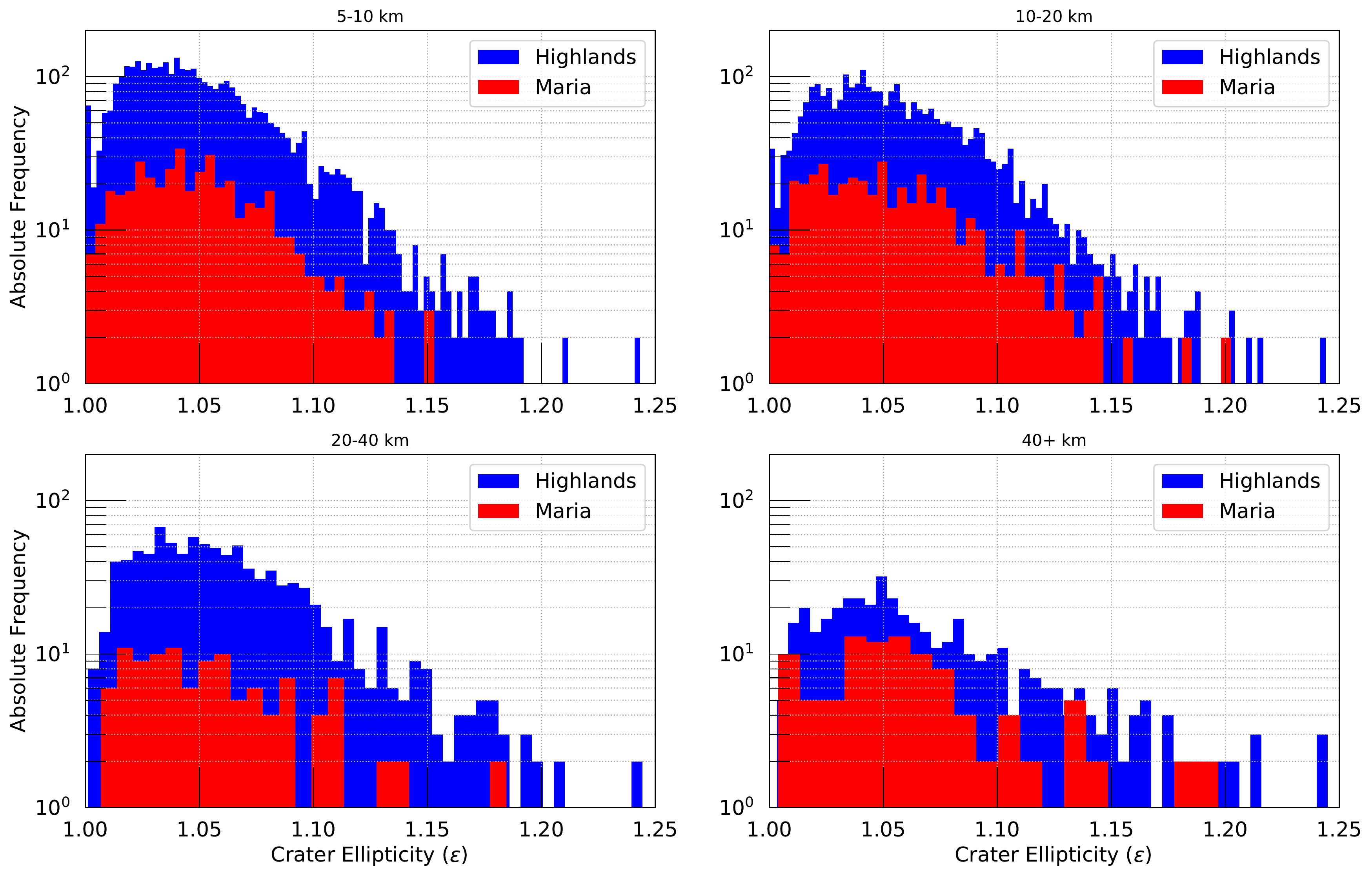}
    \caption{\textcolor{black}{Histograms of ellipticity for craters in 4 different diameter ranges, 5-10~km, 10-20~km, 20-40~km and $>$40~km.  The distributions for highlands craters are shown in blue, while the maria craters are shown in red.  Note that the bin widths in the lower two panels are twice that in the upper panels due to the smaller numbers of craters at these larger sizes.}}
    \label{fig:elldiamhist}
\end{figure}

%Finally, our model allows for the first time to search for any statistically significant differences between the highland and maria craters morphology. From visually inspecting Fig. \ref{fig:ellepticity}, it is clear that the two distributions are nearly identical. We moreover do a 2-sample Kolmogorov-Smirnov test, and obtain a K-S statistic $D=0.095$, implying that both samples were probably drawn from the same distribution. Our model hence suggest no significant ellipticity difference between highlands and maria craters.

\textcolor{black}{The importance of the strength of the target materials relative to the gravity of the target body in determining the outcome of an impact depends on the scale of the impact, with material strength being more important at smaller scales.  \citet{collins} thus predicts that there should be some variation in the ellipticity distribution as a function of crater size.  To examine this, in Fig.~\ref{fig:ellepticity2} we show the ellipticity plotted against the crater diameter.}  \textcolor{black}{Despite the large number of craters at diameters less than 20~km the proportion of craters with significant ellipticity appears lower than at larger diameters.  This is illustrated in more detail in Fig.~\ref{fig:elldiamhist}, in which we show histograms of crater ellipticity for 4 different diameter ranges.  We can see that as we move to larger diameters the distribution flattens, which indeed indicates that the proportion of elliptical craters increases with increasing crater diameter.  This is different from the prediction of \citet{collins} who suggest that on the Moon the proportion of elliptical craters should increase with increasing size for diameters larger than around 80~km, but that for craters between about 10 and 80~km the proportion of elliptical craters should be roughly flat or slightly decreasing with increasing size.} {Similar results and conclusions were also found by \cite{robbins2019}.}

\subsection{Crater depth}
\label{sec:depth}
A supplementary diagnostic for our model is the predicted crater depth distribution. This is possible to investigate since our dataset is generated from a global digital elevation map, and hence the brightness of the pixel is linearly proportional to the absolute elevation above or below a lunar reference radius. Note that LOLA's vertical precision is $\sim$ 10 cm, {but that since we downsampled the colour depth from 16-bit to 8-bit the precision of the DEM we are using is less than this}. Since ultimately what the neural network is ``seeing'' is just the pixels brightness, it is of interest to check whether the machine has a bias to certain crater depth values (and hence pixel brightness gradient). 

\begin{figure}
    \centering
    \includegraphics[width=0.8\columnwidth]{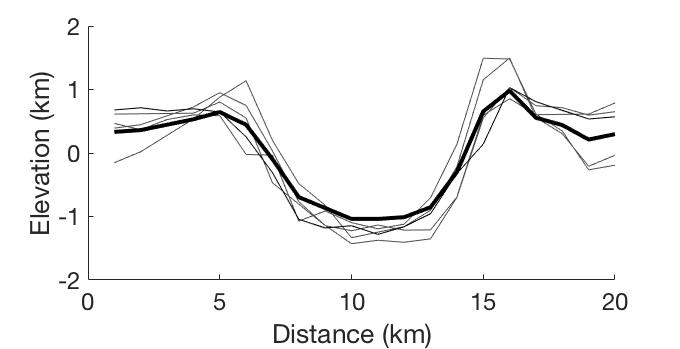}
    \includegraphics[width=0.8\columnwidth]{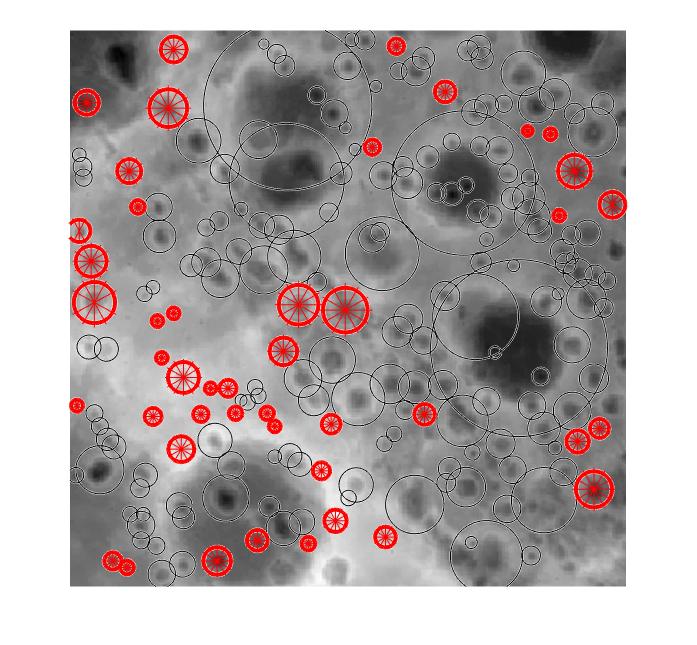}
    \caption{\textcolor{black}{Two schematics showing an example profile across a crater. In the image on the bottom, red circles show non-overlapping craters and black shows the remaining craters.
}}
    \label{fig:depthmethod}
\end{figure}

{
To calculate depth, we first start by eliminating craters that are on the edge of the DEM or overlap with another crater. For each crater we take 6 profiles at 30 degree angles that go 20\% past the crater diameter with the first profile being north-south.
We remove the background slope of the region by flattening the profiles and removing the linear trend. After averaging the profiles we calculate depth as the maximum versus minimum depth along the profile. Other techniques for calculating depth were considered, such as parabolic fits or measuring from the crater floor to the average elevation outside the crater (see \citet{robbins2018}). We found that measuring depth from the maximum to the minimum depth was the optimal method for use in automation. Our method is illustrated in Fig. \ref{fig:depthmethod}.
}

\begin{figure}
	\begin{center}
		\centerline{\includegraphics[scale=0.6]{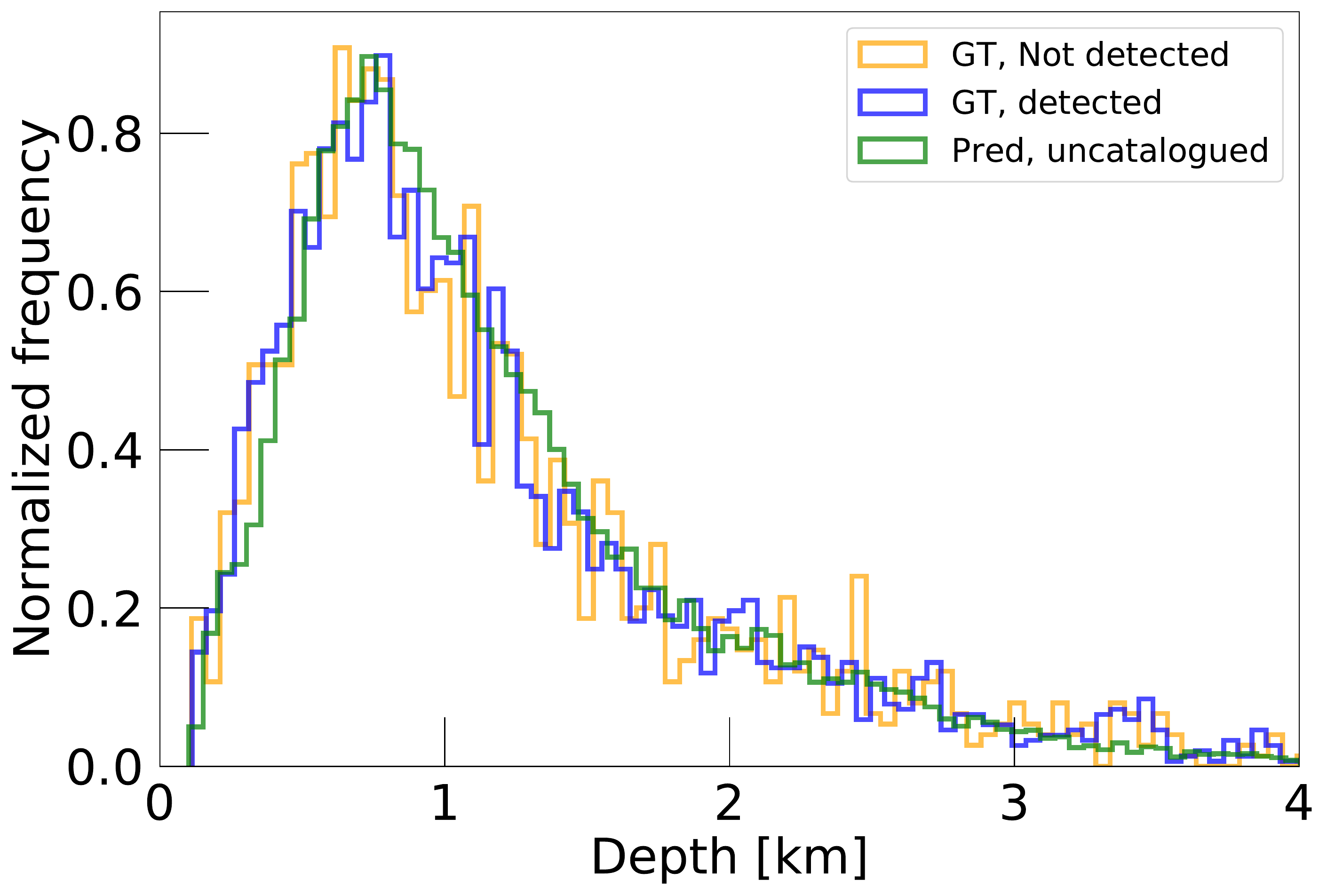}}
		\caption{Depth frequency distribution for the ground truth and predictions of our test set (post-processing). Blue represents craters that are both in GT and predicted distributions, while yellow represents craters not detected by our algorithm, and green the ``new'' craters predicted while not in the GT. }
		\label{fig:depth1}
	\end{center}
\end{figure}

%{{We construct crater depth frequency distributions using both the centroid locations and diameters taken from the ground truth database, and those predicted by our algorithm, showing both distributions in Fig. ~\ref{fig:depth1}}}
%{The ground truth and predicted craters depth frequency distributions are shown in Fig. \ref{fig:depth1}. We plot separately three distributions corresponding to craters present in both GT and predictions, craters found solely in the GT, and craters predicted but are uncatalogued in the GT.}
{To examine the performance of the machine algorithm, in Fig.~\ref{fig:depth1} we show the crater depth frequency distribution for three sets of craters, those present in both the GT and predictions, those in the GT that were not detected by the machine, and predicted craters that were not catalogued in the GT.  The depths used for craters present in the GT are those determined using the GT centres and diameters, while the depths for craters that are not catalogued in the GT are those determined using the predicted centres and diameters.  For craters present in both the GT and the predictions we find that the mean absolute deviation in the depths determined using the two sets of centres and diameters is 3.7\%, indicating that the choice of GT or predicted centre and diameter has little influence on the depth calculation {(when using this depth definition)}.  This is one reason for using the maximum to minimum depth and a fairly generous extension of the profiles beyond the crater rim as modest shifts in the crater centre and diameter should not {significantly affect} the result.}

{Looking at the histograms of the detected and undetected craters we find these two to be roughly identical, indicating that the machine does not have a significant bias toward missing craters in the existing catalogue of a specific depth. Comparing these two histograms to that of the uncatalogued craters, we find that the overall shapes of three histograms follow the same general depth distribution trends, peaking at 0.7 km. The uncatalogued craters distribution however clearly shows a slight deficiency below 0.7 km, and a slight excess beyond it. 
 This is probably due to the intrinsic difficulty in identifying shallow craters in DEMs, as they have weaker pixel intensity gradient between the rim and the center. This makes it harder for what is ultimately a crater rims identifier to spot them. There is thus a slight bias towards detecting deeper craters for those that are not in the existing catalogue, but this is not dramatic.}
{We note that crater frequency should show an increase for craters shallower than 1 km, not a peak. Our screening for non overlapping craters therefore could have introduced an additional bias.}
{Overall however, this suggests that our machine does not show significant bias with respect to depth, and is capable of detecting craters for a wide range of pixel brightness gradient, except {possibly} very shallow ones. 
In consequence, our model can be used to constrain the depth distribution of unknown craters, in addition to their sizes and shapes. }

%By examining Fig. \ref{fig:depth2}b showing the depth vs diameter distribution of craters (explained below), we can associate craters shallower than 3 km to diameters less than 15 km, implying that the excess of craters in Fig. \ref{fig:depth1} is due to the ``new'' craters not present in the ground truth (discussed in section \ref{sec:diam}). 

%The absence of an excess around 0.5 km however is puzzling. 
%While it is conceivable that our machine is less efficient at detecting craters of this depth, this seems unlikely due to the presence of the shallower craters excess (hence with weaker pixel intensity gradients). We therefore prefer the alternative interpretation that the difference in the distributions shape is physical, and is due to the machine detecting a large number of flooded, eroded, or otherwise significantly altered craters not present in the GT. These are likely to have a different depth distribution than pristine craters, and the predicted distribution would be the sum of the two. 

%Comparing detection biases as function of depth and crater flooding level is a major undertaking that we leave for future work.   

\subsubsection{Depth-diameter ratio}
\label{sec:depthdiam}

{While depth histograms are a useful diagnosis for the machine's performance, what is more meaningful physically is the depth to diameter ratio that can provide insight into strength of the target material and the crater formation process (see, for example, the review of \cite{robbins2018}). Figure \ref{fig:depth2} (top) shows our measurements for the depth versus diameter of craters in lunar highlands and maria. Shallow degraded and flooded craters are included in our data set (to the extent they are detected by the machine), creating a wide range in depth for a particular diameter. Despite this, a clear transition in slope is observable near 18 km in diameter, corresponding with the transition from simple bowl shaped craters, to complex craters \citep{pike1977}. Previous studies have found that simple craters have a depth to diameter slope of 0.1 to 0.2, while larger complex craters have have a depth to diameter slope of 0.05-0.005 \citep[e.g.][]{pike1977,stopar2017}.  Other studies such as \citet{williams1998,baker2012,daubar2014} have investigated craters that are largely outside the size ranges we consider here.  The most comparable study in number of craters, that of \citep{salamuniccar2012}, which also used a computer-aided database, estimated the diameter of the simple-complex transition but did not explicitly fit the slopes in either regime. After filtering out degraded and flooded craters, we find a similar relationship (Fig. \ref{fig:depth2} (bottom)).}

{

To identify the deepest, presumably least modified craters, we took the following steps:

\begin{enumerate}
\item Assume craters shallower than 0.5 km deep are degraded.
\item For every crater diameter, determine the average and standard deviation of crater depths in that diameter range, given a 2 km wide diameter bin.
\item Assume least modified craters are deeper than one standard deviation above the average in that size range.
\end{enumerate}

\begin{figure}
	\begin{center}
		\centerline{\includegraphics[width=\columnwidth]{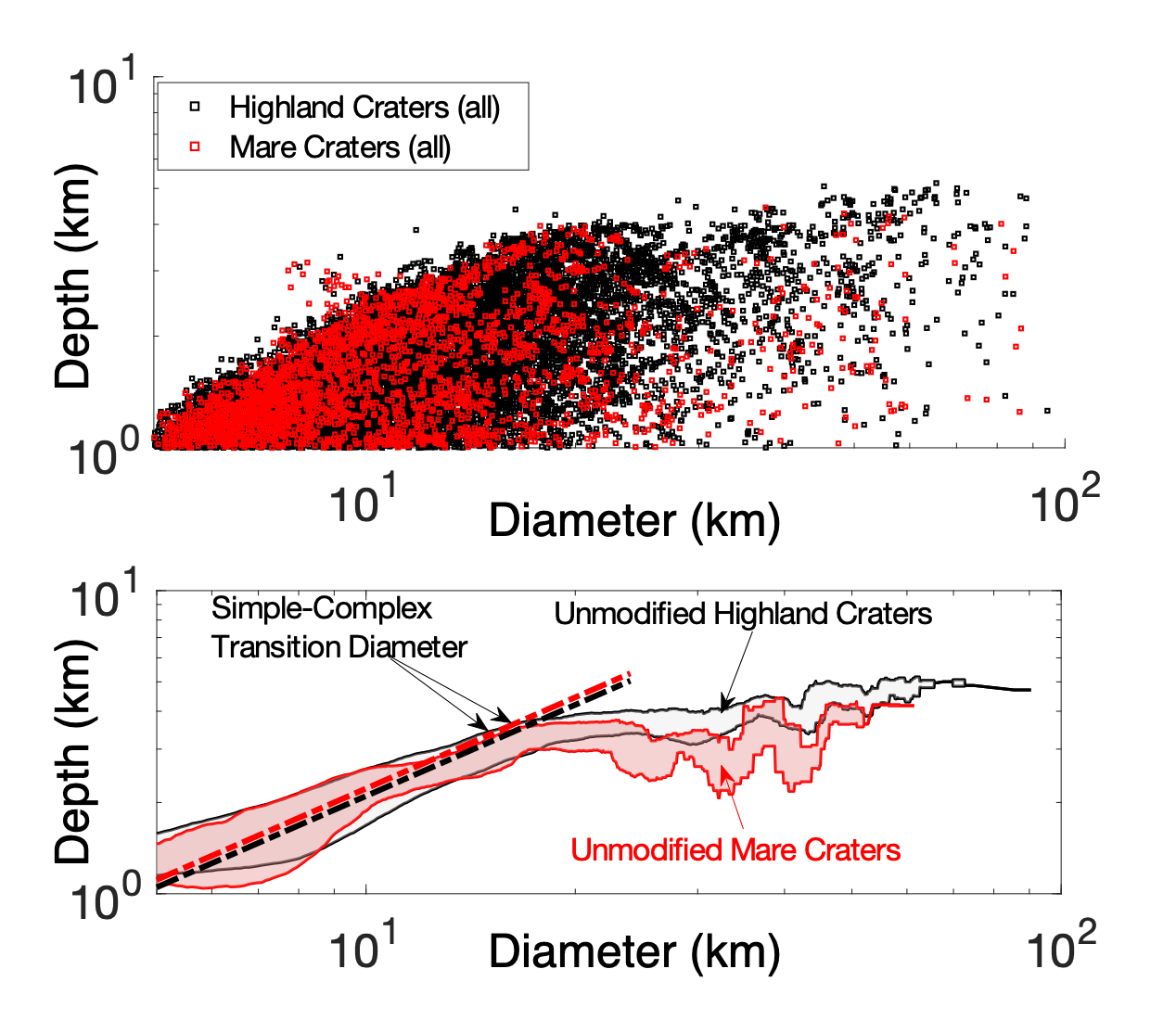}}
		\caption{Depth-diameter diagram for the predicted craters, for highlands and maria separately. The knee at $\sim$20 km diameters is due to the transition from simple to complex craters. Top panel shows all of the craters, while bottom panel shows only craters we deemed pristine. {The dashed lines are fits to the depth-diameter ratio of craters $<$15~km in diameter.  The coloured bands represent the 1-$\sigma$ spread around the mean for the unmodified craters only.}}
		\label{fig:depth2}
	\end{center}
\end{figure}

Our approach only uses elevation information to identify degraded craters as opposed to other geologic indicators that could be seen in imagery, such as evidence of craters being modified by ejecta blankets or basalt flows. However, we are confident in our assumption that the deepest craters for a given diameter should represent the least modified craters. {Implicitly we are hence assuming that these craters are not anomalous \citep{Chandnani}.} Finally, to identify the transition diameter from simple to complex craters, we perform the following steps:

 \begin{enumerate}
\item Fit depth-diameter line to the least modified craters below 15 km for both Mare and Highlands.
\item Determine the diameter when the trend breaks. 
\item Approximate this diameter and the simple to complex transition point.
\end{enumerate}
{Noting that again we do not account for anomalous craters, even though some are clearly present (for example around 8 km). However, since the goal is simply to understand the machine's performance, this simplification should be acceptable.}
It has also been predicted that the simple to complex crater transition occurs at different diameters in Mare (16 km) versus Highlands (21 km), due to differences in impact target strength \citep[e.g.][]{pike1980}. Our depth to diameter measurements suggest {transition diameter of $\sim$ 16 km in maria and $\sim$ 18 km in Highlands} consistent with \citep{pike1980,salamuniccar2012}. We also find that complex mare craters are on average shallower than complex highland craters, though this may be due to infilling by mare basalts, since as noted our purely elevation based criterion for `unmodified' craters does not result in a perfect separation.  Note that the apparent waviness in the complex craters in the lower panel of figure~\ref{fig:depth2}, especially for the maria, is due to small number statistics.  It is for this reason that we only fit the slope of the simple craters and the location of the simple-complex transition, which are not affected by small number problems.
}

%\subsection{Depth} 

%We calculate the physical depth for ground truth craters as follows:
%\begin{enumerate}
%    \item Retrieve the pixel coordinates of the ground truth mask we generated
%    \item Calculate the ``crater brightness'':  median brightness of the (local) DEM %image pixels with the same coordinates as the crater.
%    \item Calculate the ``image brightness'': mean brightness of all (local) DEM image %pixels with values larger than ``crater brightness''.
%    \item Crater depth is then defined as 0.5 $\times$ (image brightness - crater %brightness), where the 0.5 is the brightness to meters conversion factor used by LRO %LOLA.
%\end{enumerate}

%Figure \ref{fig:depth}

\section{Summary \& conclusions}
\label{conc}
In this work we trained the general computer vision framework MaskRCNN to detect lunar craters in digital elevation maps, and extract their sizes and shapes accurately. Our model was able to detect on average 87\% of all {catalogued} craters in an image, {and identify $\sim 40\%$ more craters than present in the P+H catalogs used for training and validation.} This method was moreover able to recover the observed size distribution of the test set at all diameters, improving over our previous DeepMoon algorithm for large craters, {but detecting less craters in the intermediate $\sim$ 20 km range.  This may be indicative of either a greater sensitivity to changes in crater style around the simple-complex transition for MaskRCNN than for DeepMoon, or a bias introduced by the inherent deficiency of the P+H catalogues in craters around this size.}

{We note that our machine algorithm was trained on a relatively conservative ground truth (consisting of the merged datasets of \citet{povilaitis2017} and \citet{head2010}).  Comparison of the craters identified by our machine algorithm with the database of \citet{robbins2019} suggests that our algorithm is also fairly conservative, albeit less so than the input datasets.}

One of the main advantages of MaskRCNN was retrieving the crater shapes (through the instance masks) for free, without prior explicit training. Our retrieved shape distribution {suggests that $\sim$3\% of craters larger than 5~km in diameter have ellipticities of greater than 1.2.  In terms of threshold angles we find that an impact at less than 27$^\circ$ should result in an ellipticity of greater than 1.1, while an impact at less than 10$^\circ$ should result in an ellipticity of greater than 1.2.  These results are consistent with the previous observational work of \citet{bottke} on the lunar maria, and with numerical investigations by \citet{collins}.  Our results do not match with the raw ellipticities of the database compiled by \citet{robbins2019}, which are also discordant with \citet{bottke}, however we find that when craters with less complete arcs used for the ellipse fits are excluded from \citet{robbins2019} the agreement is much better.  While our broad fractions and thresholds are largely in agreement with \citet{collins} we find that the proportion of elliptical craters increases continuously with increasing crater diameter above 5~km, which conflicts with their modelling work.}

The efficiency of our approach allowed us to search for statistical differences in the size and shape distributions of highlands and maria craters, but we found these to be roughly identical. {Automated shape retrieval has allowed us to construct one of the largest ellipticity datasets available with minimal additional cost.} {We finally found our model to perform well at all crater depths, except shallow craters $\leq$ 1.0 km deep.}

\section*{Acknowledgements}
M.A.-D. thanks the Department of Physics at the American University of Beirut where he spent one semester as a visiting faculty. The authors thank Ari Silburt for useful discussions on DeepMoon. The authors thank S. Robbins and an anonymous referee for their thorough reviews which have helped improve this manuscript.

\appendix
\section{}
\label{aa}
We adopted the following Matterport MaskRCNN model parameters in our work:\\
IMAGES$\_$PER$\_$GPU = 2 \\
IMAGE$\_$MIN$\_$DIM = 256\\
IMAGE$\_$MAX$\_$DIM = 512\\
RPN$\_$ANCHOR$\_$SCALES = (4, 8, 16, 32, 64)\\
TRAIN$\_$ROIS$\_$PER$\_$IMAGE = 600\\
RPN$\_$NMS$\_$THRESHOLD = 0.7\\
MEAN$\_$PIXEL = [165.32, 165.32, 165.32]\\
LEARNING$\_$RATE = 1e-3\\
USE$\_$MINI$\_$MASK = True\\
MAX$\_$GT$\_$INSTANCES = 400\\
DETECTION$\_$MAX$\_$INSTANCES = 400

\section{}
\label{ab}
Our \textcolor{black}{Python} preprocessing routine is provided here for reproductibility. It is applied to three identical channels in the present case.\\
\newline
\texttt{import cv2} \\
\texttt{grid$\_$size = 8}\\
\texttt{def rgb$\_$clahe$\_$justl(in$\_$rgb$\_$img):} \\
\indent \indent \texttt{bgr = in$\_$rgb$\_$img[:,:,[2,1,0]]  \textit{$\#$ switch R and B (RGB $\rightarrow $ BGR)}}\\
\indent \indent \texttt{lab = cv2.cvtColor(bgr, cv2.COLOR$\_$BGR2LAB)}\\
\indent \indent \texttt{clahe = cv2.createCLAHE(clipLimit=2.0, tileGridSize=(grid$\_$size,grid$\_$size))}\\
\indent \indent \texttt{return clahe.apply(lab[:,:,0])}\\
    

%\end{linenumbers}


  \bibliographystyle{elsarticle-harv}
  \bibliography{craters.bib}
  \begin{comment}

\end{comment}


\end{document}